\documentclass[fleqn,usenatbib,useAMS]{mnras}

\usepackage{graphicx}	
\usepackage{amsmath}	
\usepackage{amssymb}	
\usepackage{multicol}        
\usepackage{bm}		
\usepackage{pdflscape}	

\newcommand{\kms}{\,km\,s$^{-1}$} 

\def\lesssim{\mathrel{\hbox{\rlap{\hbox{\lower4pt\hbox{$\sim$}}}\hbox{$<$}}}}
\def\gtrsim{\mathrel{\hbox{\rlap{\hbox{\lower4pt\hbox{$\sim$}}}\hbox{$>$}}}}
\def\apj{ApJ}
\def\apjs{ApJS}
\def\apjl{ApJL}
\def\aj{AJ}
\def\aap{A\&\hskip-1pt A}

\def\mnras{MNRAS}
\def\pasp{PASP}
\def\araa{ARA\&\hskip-1pt A}

\def\jkas{JKAS}
\def\aapr{A\&\hskip-1pt ARv}
\def\an{AN}

\defcitealias{ann15}{ANN15}

\usepackage[T1]{fontenc}
\usepackage{ae,aecompl}

\usepackage{newtxtext,newtxmath}

\title[Dwarf Elliptical-like galaxies]{Luminosity Distribution of Dwarf Elliptical-like Galaxies}

\author[M. Seo and H. B. Ann]{
Mira Seo,$^{1}$\thanks{E-mail: mrseo@pusan.ac.kr}
Hong Bae Ann,$^{1}$
\\
$^{1}$Department of Earth Sciences, Pusan National University, 46241, Busan, Korea\\
}
\date{Accepted 2022 June 10. Received 2022 May 29; in original form 2021 October 2}

\pubyear{2021}

\begin{document}
\label{firstpage}
\pagerange{\pageref{firstpage}--\pageref{lastpage}}
\maketitle

\begin{abstract}
We present the structural parameters of $\sim910$ dwarf elliptical-like galaxies in the local universe ($z\lesssim0.01$) derived from the $r-$band images of the Sloan Digital Sky Survey (SDSS). We examine the dependence of structural parameters on the morphological types (dS0, dE,dE$_{bc}$, dSph, and dE$_{blue}$). There is a significant difference in the structural parameters among the 
five sub-types if we properly treat the light excess due to nucleation in dSph and dE galaxies. The mean surface brightness within the effective radius ($<\mu_{e}>$) of dSph galaxies is also clearly different from that of other sub-types. The frequency of disk features such as spiral arms depends on the morphology of dwarf galaxies. The most pronounced difference between dSph galaxies and other sub-types of early-type dwarf galaxies is the absence of disk feature which is thought to be closely related to their origin.
\end{abstract}




\section{Introduction} \label{sec1:intro}
Dwarf galaxies are most abundant in the local universe and they are supposed  to be building blocks of massive galaxies in the $\Lambda$CDM cosmology \citep{bos14}.  Dwarf galaxies can be divided into two groups. One is the dwarf elliptical-like galaxies (hereafter  dE-like galaxies) of which basic morphology resembles the dwarf elliptical galaxy (dE) and the other is the dwarf irregular galaxy (dI). \citet*[hereafter \citetalias{ann15}]{ann15} classified the dE-like galaxies by five subtypes, dS0, dE, dE$_{bc}
$, dSph, and dE$_{blue}$ using colour images of the Sloan Digital Sky Survey (SDSS).  The dS0 is dwarf lenticular galaxy introduced by \citet{san84} and the subtypes dE$_{bc}$ and dE$_{blue}$ represents dE galaxies with blue core and globally blue dwarfs with ellipsoidal shape, respectively.  In the traditional photographic plates, dE$_{bc}$ and dE$_{blue}$ could not be distinguished from dE. In the literature, dSph galaxies are not always distinguished from dE galaxies \citep{san84, fer94, kor12}. However, many studies treated the dSph galaxy as a distinct class of galaxy \citep{mat98, gal94, gre97, vdb99}. Among the five 
sub-types of dE-like galaxies, dS0, dE and dSph are considered to be early-type dwarfs while dE$_{blue}$ galaxies are taken as late-type galaxies if we group them by the presence of on-going star formation. The subtype dE$_{bc}$ is intermediate between them because of the active star formation in the core only. We considered dE$_{bc}$ galaxies as early-type dwarfs because their global photometric properties are similar to dS0, dE and dSph galaxies \citep{ann17}.

The dE-like galaxies have surface brightness distribution different from the  giant elliptical galaxies which are well described by the $r^{1/4}$-law. The surface brightness profile of dE-like galaxies, dE in particular, is best fitted with S\'{e}rsic index $n$ less than 2 \citep{bj98, ryd99,grant05, kim06, jan14}. Moreover, a significant fraction of dE-like galaxies show disk features which have a variety of structures such as spiral arms and bars \citep{jer00, bar02, geh03, gra03, rij03, lis06}. Most disk features can be seen from the residual images made by subtracting  axisymmetric component or from the unsharp masked images. Some of the spiral arms are grand-design arms which are mostly observed in early-type spiral galaxies \citep{elm82, ann13}.  

Presence of disk structures such as spiral arms and bars in dE galaxies raises doubts on the conventional belief that they are primordial object  collapsed in the early epoch of galaxy formation and suggests a different channel to their formation at least for dE galaxies with disk features. The most popular scenario for the origin of the early-type dwarf galaxies with disk features is that they are transformed objects from the late-type galaxies of which their cold gas was removed by the ram pressure from the hot intracluster medium \citep{gg72}. In addition, the structure of a galaxy can be 
modified by the tidal field of the group and cluster when they enter into them  \citep{moo98, moo99}. The rapid rotation observed in some dwarf elliptical galaxies supports the hypothesis that the dwarf elliptical galaxies are transformed from the spiral galaxies by losing their gas in groups and clusters. Since the majority of dE-like galaxies are members of groups or clusters \citep{ann17}, the environmental quenching \citep{pen10}, a sudden shutdown of star formation, caused by ram pressure stripping \citep{gg72} together with galaxy harassment \citep{moo98, moo99} and tidal shock \citep{may01} can transform late-type galaxies into early-type dwarfs.

It seems of interest to see whether the structural parameters and the embedded disk features of dE-like galaxies are distinguished for the five sub-types. 
In particular, we are very interested to see whether the structural difference between dE and dSph is significant enough to reflect their morphological difference. In general, dE-like galaxies with $M_{B} < -14$ are considered as dwarf elliptical galaxies while those fainter than $M_{B} = -14$ are considered as dwarf spheroidal galaxies. However, there are some overlaps in their luminosity distribution  \citepalias{ann15}. Moreover, the morphological difference between dE and dSph is based on the differences in surface brightness and its gradient \citepalias{ann15}. The dSph galaxies show faint surface brightness with shallow gradient than the dE galaxies. In addition, there is pronounced differences in the dynamical property and dark matter content. The dE galaxies are supposed to be supported by the rapid rotation while the dSph galaxies are supported by the velocity dispersion \citep{geh02,geh03,koc07}. However, it does not mean 
that all dSph galaxies have no rotation and all dE galaxies are fast rotators. Rather, some dSph galaxies have significant rotations \citep{del17} and some dE galaxies are slow rotators \citep{tol15}. The dSph galaxies have much larger fraction of dark mater than the dE galaxies \citep{mat98, gil07}.

To examine the structural differences among dE-like galaxies, we analysed their luminosity distributions using the GALFIT \citep{pen02} which is known to be effective to analyse the two-dimensional luminosity distribution of galaxies. We examined the residual images obtained by subtracting the GALFIT model images from the observations to see the disk features. Since the majority of dS0, dE, and dSph galaxies have nuclear component \citepalias{ann15}, we apply multi-component models to dS0, dE, dSph galaxies along with dE$_{bc}$ galaxies. The main purpose of the present study is to derive the S\'{e}rsic index of the main component to know their statistics. We also examine whether the sub-types of the dE-like galaxies have significantly different structures. 

The present paper is organized as follows. In section II, we describe the data and method. The results are presented in section III. Discussion is presented in section IV and a brief summary and conclusions are given in the last section.

\section{Data and method} \label{sec2:data and method}
\subsection{Data selection} \label{subsec1:data}
\begin{figure}
\graphicspath{ {./figs/} }
\centering
\includegraphics[width=\columnwidth]{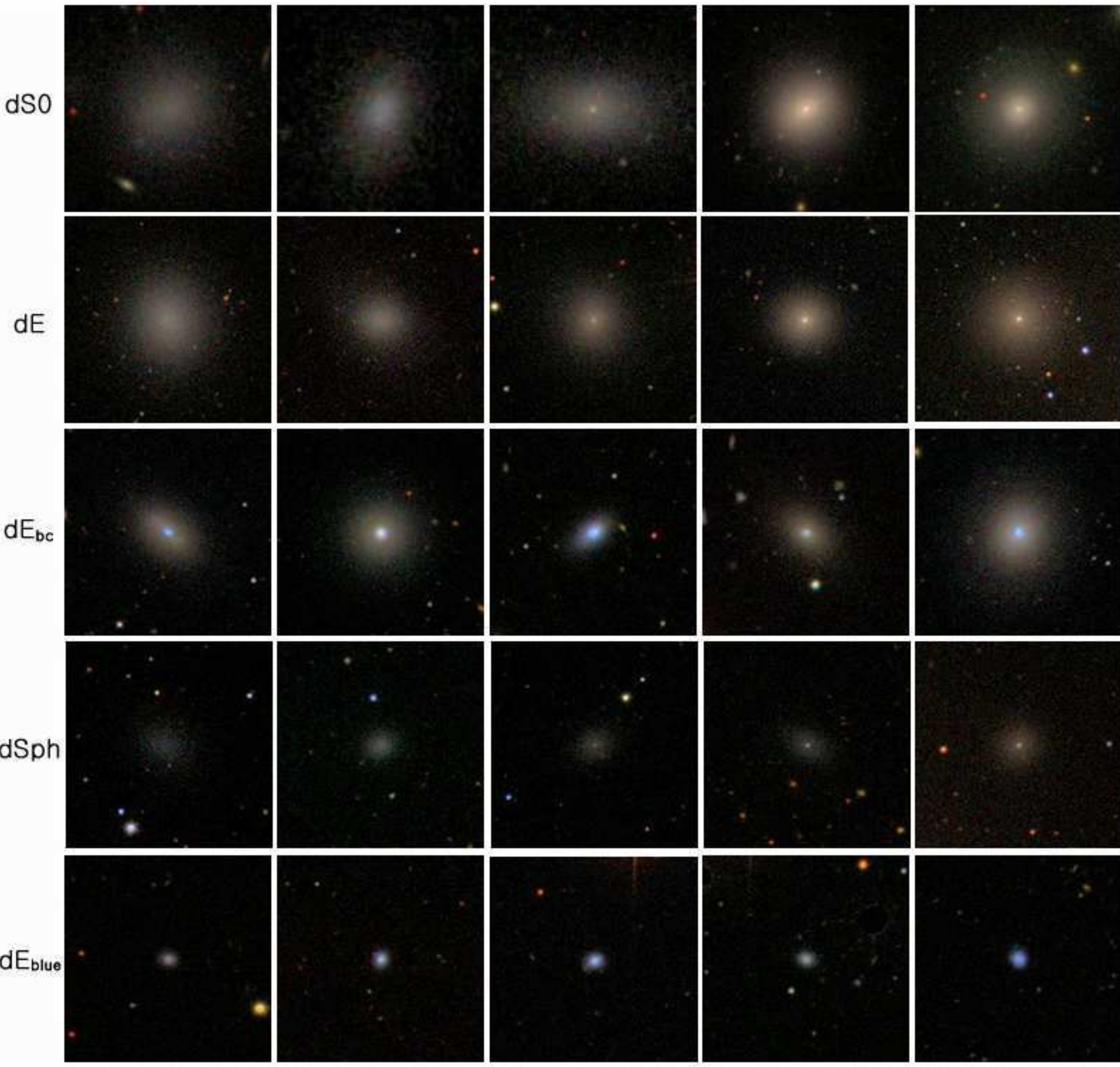}
\caption{Sample images for dwarf elliptical-like galaxies: dS0 (first row), dE (second row), dE$_{bc}$ (third row), dSph (fourth row) and dE$_{blue}$ (fifth row). Left two coloumns are non-nucleated galaxies and right three coloumns are nucleated galaxies in dS0, dE, dSph galaxies. (It is made by  the images in Figure 5 of \citetalias{ann15}).
}
\label{fig1}
\end{figure}

Our galaxy sample is $\sim$910 dE-like galaxies divided into 5 sub-types (dS0, dE, dSph, dE$_{bc}$, dE$_{blue}$) from the galaxy catalogue of  \citetalias{ann15}. It is a visually classified galaxy catalogue of the local universe with z $\lesssim$ 0.01 using SDSS DR7. Our dE-like galaxy sample is a part of the 5,386 galaxies of  \citetalias{ann15}.  
Some sample galaxies are presented in Figure \ref{fig1}.

Data of \citetalias{ann15} is basically taken from KIAS-VAGC \citep{chk10} and some faint galaxies are added from the NED(NASA Extragalactic Data) and \citet{mk11} catalogues.  KIAS-VAGC is based on the New York University Value-added catalogue(NYU-VAGC; \citet{aba09}) using SDSS DR7 and 
contains several physical quantities of galaxies including the computer-based automatic classification \citep{park05}. \citet{mk11} catalogue is a list of galaxy group in the local universe with ${\left| b \right| > 15\degr}$ and with $V_{LG} < 3,500$\kms for a local Group, and have classifications of 10,914 galaxies based on the RC3 classification. 

We used the galaxy distances given by \citetalias{ann15} who calculated the distances of galaxies using the V$_{LG}$ obtained from the observed redshift by the method of \citet{mou00} assuming the Hubble constant $H = 75$ \kms\,Mpc$^{-1}$. For the galaxies in the Virgo Cluster, the distance of the Virgo Cluster was applied. On the other hand, for galaxies of which the metric distances are given in the NED, the metric distances were used. 

\subsection{Method} \label{subsec2:method} 
We use r-band images of SDSS \citep{york2000,sto02} Data Release 7 \citep{aba09} 
for GALFIT \citep{pen02, pen10}. We first constructed a single-component model with a S\'{e}rsic function which is known to represent the surface brightness distribution of dE-like galaxies. In the case of dS0 galaxies, which are early-type dwarf galaxies containing disk component \citep{BST85}, so we have applied two-components model with a S\'{e}rsic $+$ exponential disk functions. And the basic parameters like integrated magnitude, effective radius, axis ratio and position angle, required for GALFIT model are taken from the SDSS casjob. However, applying a single component model to nucleated dE, dS0 and dSph galaxies, there is an excess of light in the central region of the residual image, which appears as a ring, oval, or disk in the central regions of galaxies. 

\begin{table} 
  \centering
  \caption{Fitting models of dE-like galaxies for GALFIT}
    \begin{tabular}{|l|l|}
    \hline
        $type$     & Functions \\
    \hline
    dS0$_{n}$  & S\'{e}rsic $+$ expDisk $+$ King$/$Nuker \\
    dS0$_{un}$  & S\'{e}rsic $+$ expDisk  \\
    dE$_{n}$  & S\'{e}rsic $+$ King$/$Nuker \\
    dE$_{un}$ & S\'{e}rsic  \\
    dSph$_{n}$  & S\'{e}rsic $+$ King$/$Nuker \\
    dSph$_{un}$ & S\'{e}rsic \\
    dE$_{bc}$  & S\'{e}rsic $+$ King$/$Nuker \\
    dE$_{blue}$ & S\'{e}rsic \\
    \hline
    \end{tabular}%
    \label{tab1}%
\end{table}%

To treat the excess of light in the central regions of galaxies, we add a nuclear component to single  S\'{e}rsic function. There are two useful functions in GALFIT for fitting the luminosity of the nuclear region of dwarf galaxy, the empirical (modified) King profile\citep{els99} and the Nuker profile\citep{lau95}.  We apply these functions to analyse the luminosity of our sample. Thus, we have modelled the nucleated dE-like galaxies of dE$_{n}$, dSph$_{n}$, and dE$_{bc}$ with the S\'{e}rsic profile $+$ King$/$Nuker profile, and non-nucleated galaxies of dE$_{un}$, dSph$_{un}$, and dE$_{blue}$ with the S\'{e}rsic profile. And for dS0 galaxies,  we used  three and two components models including exponential disk profile for dS0$_{n}$ and dS0$_{un}$ galaxies, respectively (see Table \ref{tab1}).

\section{Results}\label{sec:results}
We examined the structural parameters derived from the GALFIT along with the luminosity ($M_{r}$) derived from the model $r-$magnitude corrected for the extinction caused by the dust in the Galaxy. We described the S\'{e}rsic index ($n$) separately but other parameters such as $R_{e}$ and <$\mu_{e}$> are examined in the correlation analysis only. Here, $R_{e}$ is the effective radius within which half of the galaxy’s luminosity is contained and <$\mu_{e}$> is the mean surface brightness within $R_{e}$.

\subsection{Luminosity of dE-like galaxies} \label{subsec:luminosity}
\begin{figure}
\graphicspath{ {./figs/} }
\centering
\includegraphics[width=\columnwidth]{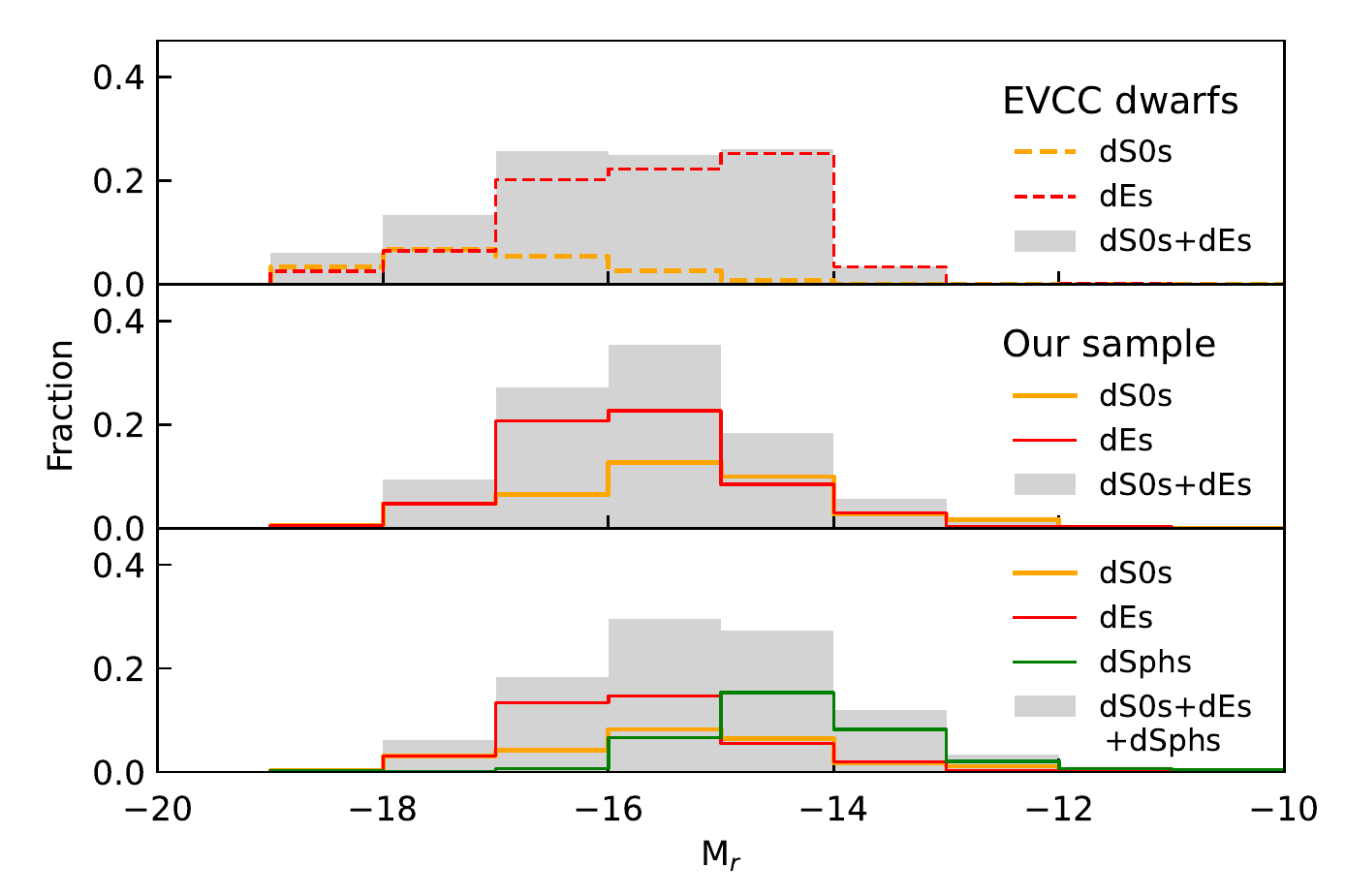}
\caption{Luminosity distributions of early type dwarf galaxies of our sample and EVCC.
Top panel is dE+dS0 galaxies of the EVCC. Middle panel shows dS0s+dEs of our sample and bottom panel presents dEs+dS0s+dSphs of our sample. The grey shaded regions show the sum of sub-types.
Red lines represents dE galaxies, orange lines for dS0s and green lines for dSphs, respectively.}
\label{fig2}
\end{figure}

Dwarf galaxies are known to have luminosity fainter than M$_{B}\approx-18$.  Figure \ref{fig2} shows the luminosity distributions of the three sub-types (dS0, dE, and dSph) of dE-like galaxies, compared with that of the dE-like galaxies in the EVCC \citep{kim14}. 
We use a distance of 16.65 Mpc for the Virgo cluster to convert the SDSS $r$-band model magnitudes of sample galaxies to the absolute magnitudes $M_{r}$. The top panel presents the luminosity distribution of the EVCC sample. The total sample is plotted as gray-coloured  region, and red and orange lines represent the distributions of dEs (red) and dS0s (orange), respectively. The middle and bottom panels show the luminosity distribution of our sample; dS0s and dEs in the middle panel and dS0s, dEs, and dSphs in the bottom panel. The green line in the bottom panel represent dSph galaxies in our sample.

At a glance, there seems to be a large difference between the luminosity distributions of dE and dS0 galaxies in the two samples. There are a lot of faint dE galaxies ($M_{r} > -15$) in the EVCC while only a small fraction of faint dE galaxies in our sample. It is due to the inclusion of dSph galaxies in the dE sample of the EVCC, whereas we separate dSph galaxies from the dE sample. Almost all dSph galaxies in our sample are classified as dE in the EVCC. As shown in the bottom panel of Figure \ref{fig2}, the luminosity distribution of dSph galaxies is much different from those of dE and dS0 galaxies. The fraction of dSph galaxies that are fainter than $M_{r} = -15$ is about three times larger than that of dSph galaxies that are brighter than $M_{r} = -15$. Our sample has more fainter dE-like galaxies than the EVCC sample because our sample includes local dwarf galaxies whose distances are closer than the Virgo cluster.

If we compare the luminosity distributions of dwarf galaxies without sub-type division, which are plotted as grey regions in the top and bottom panels, there is not much difference except for the peak luminosity. The difference in the luminosity distributions of the EVCC dwarfs (top panel) and our sample (bottom panel) is due to the difference in the mean distance of the two samples. Our sample consists of all the early-type dwarfs (dS0 + dE + dSph) with redshift less than $z=0.01$ observed in the SDSS while the EVCC sample consists of dwarfs in the Virgo cluster only. Thus, our sample includes more fainter dwarfs than the EVCC sample. 
However, it is also affected by the different environment of the two samples. The higher fraction of fainter galaxies ($M_{r} > -14$) in our sample is due to the galaxies closer than the Virgo cluster whereas the higher fraction of brighter galaxies ($M_{r} < -16$) in the EVCC is thought to be caused by the environmental effect since the high galaxy density of the Virgo cluster is favorable for luminous dwarfs if the luminosity-density relation observed for giant galaxies (e.g., \citealt{park07}) holds for dwarf galaxies.

Another point worth to be noted is that the luminosity distributions of dE and dS0 galaxies are quite similar in our sample (middle panel) while those of EVCC sample are very different. The highest fraction of dS0 in our sample occurs at $M_{r}$=-15.5 while that of EVCC sample occurs at $M_{r}$=-17.5, two magnitudes brighter. It seems to be due to classification criteria for dS0 galaxies applied by \citetalias{ann15} who considered a dE-like galaxy with a central lens-like component as a dS0 galaxy while dS0 in the EVCC sample has a lens or small bulge. We think some of the brightest dS0 in the EVCC sample are S0 galaxies in \citetalias{ann15} as suggested  by their luminosity.

\subsection{S\'{e}rsic index of dE-like galaxies} \label{subsec:sersic index}
We conducted a two-dimensional image decomposition using the $r$-band images of dE-like galaxies by applying the GALFIT \citep{pen02} in order to analyse their photometric properties. GALFIT assumes the S\'{e}rsic function as the basic luminosity profiles of galaxies together with several other functions such as King profile \citep{els99} and Nuker profile \citep{lau95}. The S\'{e}rsic profile is expressed as $$ \mu(R) = \mu_{e} + 1.086 b_{n}[({R \over R_{e}})^{1/n} -1]  $$ 
where $\mu_{e}$ is effective surface brightness at $R_{e}$, n is the S\'{e}rsic index, and $b_{n}$ is coupled to n to ensure that half of the total flux is always within $R_{e}$.

\begin{table*}
  \centering
  \caption{Fraction of S\'{e}rsic Index (n) of dE-like galaxies. }
    \begin{tabular}{|c|c|c|c|c|c|c|c|c|c|c|c|}
    \hline
        $n$     & dS0  & dS0$_{un}$ & dS0$_{n}$  & dE  & dE$_{un}$  & dE$_{n}$  & dSph & dSph$_{un}$ & dSph$_{n}$ & dE$_{bc}$  & dE$_{blue}$ \\
    \hline
    0 $\sim$ 0.5 & 0.13  & 0.09  & 0.04  & 0.00  & 0.00  & 0.00  & 0.01  & 0.00  & 0.01  & 0.01  & 0.01  \\
    0.5 $\sim$ 1.0 & 0.24  & 0.12  & 0.11  & 0.18  & 0.06  & 0.13  & 0.52  & 0.07  & 0.45  & 0.18  & 0.20  \\
    1.0 $\sim$ 1.5 & 0.32  & 0.17  & 0.15  & 0.62  & 0.07  & 0.55  & 0.27  & 0.05  & 0.22  & 0.48  & 0.46  \\
    1.5 $\sim$ 2.0 & 0.11  & 0.03  & 0.09  & 0.11  & 0.00  & 0.11  & 0.00  & 0.00  & 0.00  & 0.21  & 0.13  \\
    2.0 $\sim$ 2.5 & 0.07  & 0.02  & 0.05  & 0.03  & 0.00  & 0.02  & 0.00  & 0.00  & 0.00  & 0.07  & 0.02  \\
    2.5 $\sim$ 3.0 & 0.02  & 0.01  & 0.01  & 0.00  & 0.00  & 0.00  & 0.00  & 0.00  & 0.00  & 0.02  & 0.00  \\
    3.5 $\sim$ 4.0 & 0.00  & 0.00  & 0.00  & 0.00  & 0.00  & 0.00  & 0.00  & 0.00  & 0.00  & 0.00  & 0.00  \\
    \hline
    \end{tabular}%
    \label{tab2}%
\end{table*}%

\begin{figure}
	\graphicspath{ {./figs/} }
	\centering
	\includegraphics[trim=40 200 20 150,width=\columnwidth]{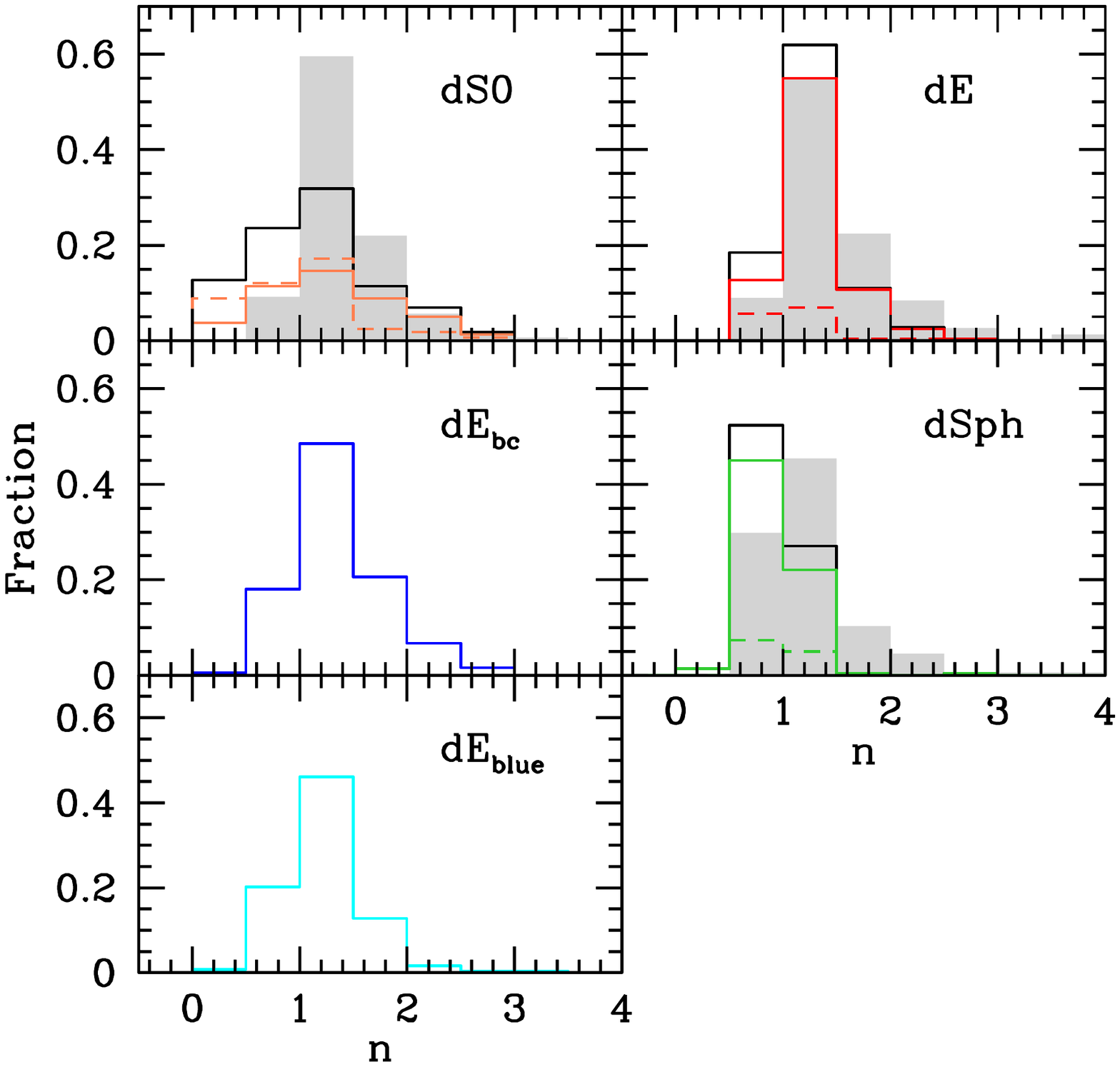}
	\caption{Frequency distributions of the S\'{e}rsic index of dE-like galaxies. The total distribution is represented by black solid line and galaxies with and without nucleation are coloured solid and dashed lines, respectively. Shaded regions designate the  S\'{e}rsic index derived from single component models.}
	\label{fig3}
\end{figure}

Figure \ref{fig3} shows the distributions of the S\'{e}rsic index (n) derived for the five sub-types of dE-like galaxies. For dS0, dE, and dSph galaxies, the distributions of S\'{e}rsic index n are further divided into the galaxies with and without nucleation. We present the total distributions by black solid lines and the galaxies with and without nucleation by colored solid and dashed lines, respectively. In cases of dE$_{bc}$ and dE$_{blue}$ galaxies, only the total distributions are plotted. The dE$_{bc}$ galaxies are considered to have no nuclei but cores and the dE$_{blue}$ galaxies have no nuclei. The S\'{e}rsic index n presented in Figure \ref{fig3} is the S\'{e}rsic index fitted to the main component of each galaxy. In cases of nucleated dS0, dE,and dSph galaxies which are denoted as dS0$_{n}$, dE$_{n}$, and dSph$_{n}$, respectively in the catalogue of \citetalias{ann15} and dE$_{bc}$ galaxies that have blue cores, we used two or three component models in GALFIT. One is the S\'{e}rsic profile and the other is either King profile or Nuker profile. For dS0 galaxies, exponential disk profile is added to other models. We considered the King profile preferentially as the luminosity distribution of a nucleation, but  we used the Nuker profile when the King profile fails to approximate the nuclear luminosity. We summarize the fitting models in Table \ref{tab1}.  

The S\'{e}rsic index n, as shown in Figure \ref{fig3}, is in the range of 0.5 to 3.0 with some differences among sub-types. When we apply a single component model (shaded region), the dS0 and dE galaxies show similar distributions. But, they are quite different if we use  S\'{e}rsic $+$ exponential disk $+$ nuclear component in dS0s and  S\'{e}rsic + nuclear component in dEs. 
The dS0 galaxies have $n$ mainly in the range of 0.5$\sim$2.0, while dE galaxies have $n$ in the range of 1.0 $\sim$2.0. The dE$_{bc}$ galaxies have similar ranges as the dE galaxies, but have more fractions in larger n. The S\'{e}rsic index n of dSph and dE$_{blue}$ galaxies are mostly in the range between 0.5 and 1.5. 
As shown in Table \ref{tab2}, more than $50\%$ of dE galaxies have $n$ between 1.0 and 1.5. The most frequent $n$ in dS0s is similar to that of dEs, but the distributions of $n$ are not similar. More than $50\%$ of dEs have $n$ between 1.0$\sim$1.5, while less than $40\%$ of dS0s have $n$ in the same range. The distribution of $n$ in dS0s is wider than that in dEs, and the fractional frequency is slightly decreased toward small $n$ in dS0s.
The dSph galaxies have a different distribution from dS0 and dE galaxies. The maximum frequency of $n$ is shifted to a smaller value, and the width of the distribution is narrow. In other words, galaxies with $n = 0.5\sim1.0$ are most abundant, followed by galaxies with $n =1.0\sim1.5$. 

We applied the Kolmogorov-Smirnov (K-S) test to examine the statistical significance of the differences in the frequency distributions of the S\'{e}rsic index n among the five sub-types of dE-like galaxies. We summarize the result of K-S test in Table \ref{tab3} where $D_{K-S}$ is K-S statistic representing the maximum difference in the cumulative distributions of two samples and p is the p-value corresponding to $D_{K-S}$. We used the K-S statistic $D_{\alpha}$ with a significance level of $\alpha$ = 0.05 to test the null hypothesis that the two samples are drawn from the same population. We rejected the null hypothesis if $D_{K-S} \geq D_{\alpha}$ or p-value less than $\alpha$. As can be inferred from the p-values in Table \ref{tab3}, each sub-type of dE-like galaxies has S\'{e}rsic index statistically different from those of other sub-types except for similar S\'{e}rsic index for dE and dE$_{blue}$ galaxies. However, there is no significant distinction between the nucleated dwarfs and non-nucleated dwarfs except for dSph galaxies.

\begin{table}

	\centering
	\caption{K-S test of S\'{e}rsic Index (n) for five sub-types of dE-like galaxies. }
	\begin{tabular}[width=\columnwidth]{|l|c|c|c|c|}
	\hline
		types    & $D_{K-S}$ & p &  sample~size  \\
		\hline
		dS0, dE                &   0.2781 &    < 0.0001 &   140, 231 \\
		dS0, dE$_{bc}$         &   0.2623 &    < 0.0001 &   140, 186 \\
		dS0, dSph              &   0.3606 &    < 0.0001 &   140, 179 \\
		dS0, dE$_{blue}$       &   0.2559 &    < 0.0001 &   140, 229 \\
		dE, dE$_{bc}$          &   0.1698 &    0.0053   &   231, 186 \\
		dE, dSph               &   0.4720 &    < 0.0001 &   231, 179 \\
		dE, dE$_{blue}$        &   0.0894 &    0.3172   &   231, 229 \\
		dE$_{bc}$, dSph        &   0.5289 &    < 0.0001 &   186, 179 \\
		dE$_{bc}$, dE$_{blue}$ &   0.1960 &    0.0008   &   186, 229 \\
		dSph, dE$_{blue}$      &   0.3958 &    < 0.0001 &   179, 229 \\
		dS0$_{n}$, dS0$_{un}$  &   0.2271 &    0.0542   &    72,  68 \\
		dE$_{n}$, dE$_{un}$           &   0.4293 &    0.0001   &   198,  33 \\
		dSph$_{n}$, dSph$_{un}$&   0.1898 &    0.3806   &   152,  27 \\
		
		\hline
	\end{tabular}%
      	
	\label{tab3}%
\end{table}%

\subsection{Correlations between structural parameters} \label{subsec:structure}

\begin{figure*}
\graphicspath{ {./figs/} }
\centering
\includegraphics[trim=20 180 30 80,width=\textwidth]{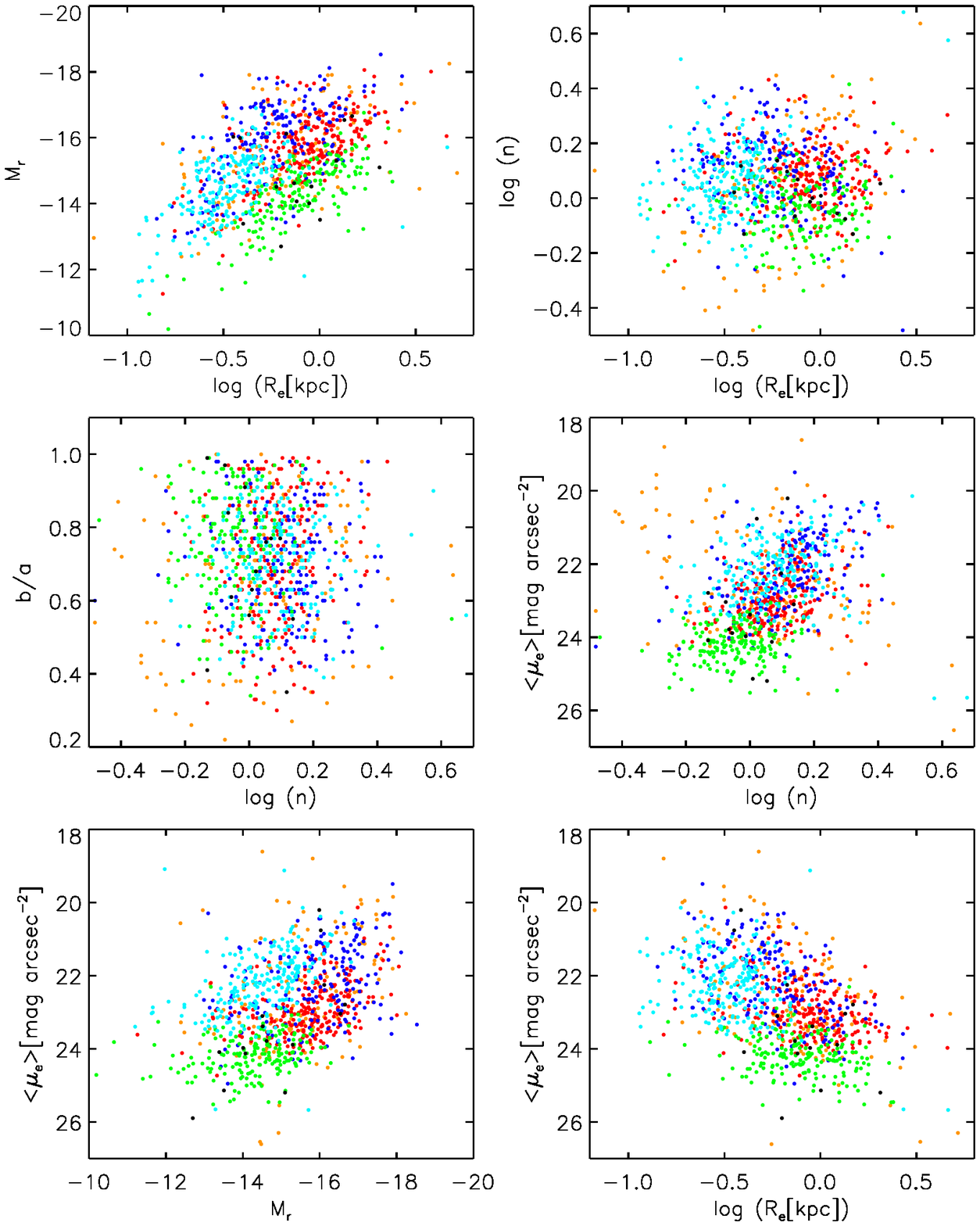}
\caption{Correlations between GALFIT parameters of dE–like
galaxies. (Top left) log ($R_{e}$) vs $M_{r}$, (top right) log ($R_{e}$) vs log($n$),
(Middle left) log ($n$) vs $b/a$, (middle right)  log ($n$) vs $<\mu_{e}>$, 
(bottom left) $M_{r}$ vs $<\mu_{e}>$, (bottom right) log $R_{e}$) vs $<\mu_{e}>$. Colour codes are 
orange(dS0), red(dE), blue(dE$_{bc}$), green(dSph), and cyan(dE$_{blue}$).
 }
\label{fig4}
\end{figure*}

We examine the correlations between the structural parameters derived from
GALFIT ($n$, $R_{e}$, $<\mu_{e}>$, $b/a$) together with $M_{r}$. Here $<\mu_{e}>$ is the mean surface brightness within $R_{e}$. We present some examples in Figure \ref{fig4}. We use colour coded symbols for easy distinction of the sub-types: dS0(orange), dE(red), dE$_{bc}$(blue), dSph(green), and dE$_{blue}$(cyan). There are fairly good correlations among the three 
parameters, $M_{r}$, $R_{e}$, and $<\mu_{e}>$ if we consider all the sub-types as a whole, whereas axial ratio ($b/a$) shows little or no correlations with other parameters. In case of the S\'{e}rsic index n, it shows a weak correlation with  $<\mu_{e}>$ if we consider all the sub-types except dS0. What is remarkable is that the sub-types of dE-like galaxies are fairly well distinguished in the diagrams for parameters $M_{r}$, $R_{e}$, and $<\mu_{e}>$ except for the indistinguishable two sub-types. 
dE and dS0, which have similar structural parameters with $M_{r} = -15.73 \pm1.11$, log ($R_{e}) = -0.09 \pm 0.28$, $<\mu_{e} > = 22.86 \pm 0.97$ for dE galaxies and $M_{r} = -15.37 \pm 1.54$, log ($R_{e}) = -0.19 \pm 0.36$, $<\mu_{e}> = 22.41\pm 1.66$ for dS0 galaxies, respectively. The reason for the similarity in the structural parameters of dS0 and dE galaxies is their similar origin, transformed from the late-type galaxies.

Among the correlations between structural parameters of dE-like galaxies, the correlation between  $<\mu_{e}>$ and $M_{r}$ presented in the bottom left panel of Figure \ref{fig4} is of a special interest because it plays a key role to distinguish dwarf ellipticals from giant ellipticals. \citet{kor85} showed that the correlation between  $<\mu_{e}>$ and $M_{r}$ for dwarf elliptical galaxies shows a trend opposite to the correlation observed for giant elliptical galaxies. There is a count argument that the two relations are continuous \citep{gra03}. Detailed analysis of the correlation between $<\mu_{e}>$ and $M_{r}$ were given by \citet{kor12} which supports the opposite behaviour raised by \citet{kor85}. It dictates the increasing luminosity with increasing surface brightness for dwarf elliptical galaxies and an opposite trend for giant elliptical galaxies. 

As shown in Figure \ref{fig4}, most correlations show weak or moderate correlations with the strongest one for the correlation between $M_{r}$ and log ($R_{e}$) which has the Pearson correlation coefficient (cc) of  -0.58. However, if we examine the correlation between $M_{r}$ and log ($R_{e}$) for each sub-type of dE-like galaxies, the correlation strengths become higher with the strongest correlation for dSph galaxies that have cc = -0.71 and the weakest correlation for dE$_{bc}$ galaxies that have cc = -0.59. The dS0, dE and dE$_{blue}$ galaxies have similar correlation strength as cc = -0.63, -0.67, and -0.65, respectively. The slope, i.e., regression coefficient, of the relation between  $M_{r}$ and log ($R_{e}$) for dSph galaxies is $-3.18 \pm 0.23$ with confidence interval of (-4.53, -1.85) and the slopes of other sub-types are within the confidence interval for dSph which is derived from the 1000 bootstrap resampling. The strongest correlation observed for the structural parameters $M_{r}$ and $R_{e}$ seems to reflect the well established relationship between the luminosity and size of galaxies \citep{bla03, she03}.   

It is clear that the luminosity of dSph galaxies is similar to dE$_{blue}$ galaxies but they have 
completely different surface brightness. On average, the surface brightness of dSph galaxies 
is lower than other sub-types of dE-like galaxies whereas that of dE$_{blue}$ galaxies is 
higher than other sub-types except for dE$_{bc}$ galaxies of which surface brightness is similar 
to dE$_{blue}$ galaxies. The high surface brightness of dE$_{blue}$ galaxies is due to their 
small size. Since dSph galaxies share structural properties with dS0, dE, and dE$_{bc}$ 
galaxies except for $<\mu_{e}>$, they do not share structural properties with dE$_{blue}$ 
galaxies except for the luminosity and axis ratio ($b/a$). It seems likely that formation and
evolution of dSph and dE$_{blue}$ galaxies are much  different. On the other hand, dS0 galaxies 
are difficult to distinguish from other sub-types of galaxies because bright part of these galaxies 
($M_{r}$  < -15) are almost indistinguishable in surface brightness and luminosity from dE galaxies,
and faint part lies in the regions where dE$_{blue}$ galaxies are likely to be located. The dE$_{bc}$ 
galaxies are located everywhere except for the regions with $M_{r}$ > -13. The dE galaxies, 
which are a typical type of dE-like galaxies, are distinct from dSph or dE$_{blue}$ galaxies, 
but they have structural parameters similar to dS0 and dE$_{bc}$ galaxies.

As shown in the bottom right panel of Figure \ref{fig4} there is a moderate correlation between 
$<\mu_{e}>$ and log $(R_{e})$ of dE-like galaxies with cc = 0.46. The effective radii decrease 
with increasing surface brightness. A similar correlation was reported for giant elliptical galaxies 
\citep{kor85} with much smaller dispersion. The largest source of dispersion is the dS0 
galaxies which have regression coefficient, i.e., slope, of $1.67 \pm 0.36$ of which the error of the regression coefficient is about two times larger than those of other sub-types. This makes the correlation strength (cc=0.37) weaker than other sub-types. Among the five sub-types, dSph galaxies have the smallest slope ($1.04 \pm 0.16$) which is manifested by their shallow surface brightness gradient. The dSph and dE$_{blue}$ galaxies are quite well segregated in the $M_{r}$ $vs$ log ($R_{e}$) and $<\mu_{e}>$ $vs$ $M_{r}$ diagrams as well as $<\mu_{e}>$ $vs$ log $(R_{e})$ diagram.


As mentioned above, the S\'{e}rsic index n shows weak dependence on other structural 
parameters. As shown in the middle right panel of Figure \ref{fig4}, dE-like dwarfs show decreasing $<\mu_{e}>$ with increasing log (n), which have the regression coefficient of $-0.93 \pm 0.21$ with cc = -0.14, for the whole sample. In contrast, if we examine the correlation between $<\mu_{e}>$ and log $(n)$ by considering the sub-types, there are large differences among sub-types. For instance, dSph galaxies show nearly no correlation between $<\mu_{e}>$ and log (n), i.e., the regression coefficient of $0.53 \pm 0.13$ with cc = 0.29 and opposite trend for dS0 galaxies that have the regression coefficient of $0.65 \pm 0.19$ with cc = 0.29. The negative correlation of the dE-like galaxies is mainly due to the negative regression coefficients of dE, dE$_{bc}$, and dE$_{blue}$ galaxies among which dE$_{bc}$ galaxies have the strongest correlation with cc = -0.41. Because a large $n$ indicates a steep luminosity gradient which results in high surface brightness in the inner part of a galaxy, the negative regression coefficients found for dE, dE$_{bc}$, and dE$_{blue}$ galaxies seem to be natural.  We present the results of correlation analysis among the structural parameters in the Appendix.

\begin{figure}
\graphicspath{ {./figs/} }
\centering
\includegraphics[width=\columnwidth]{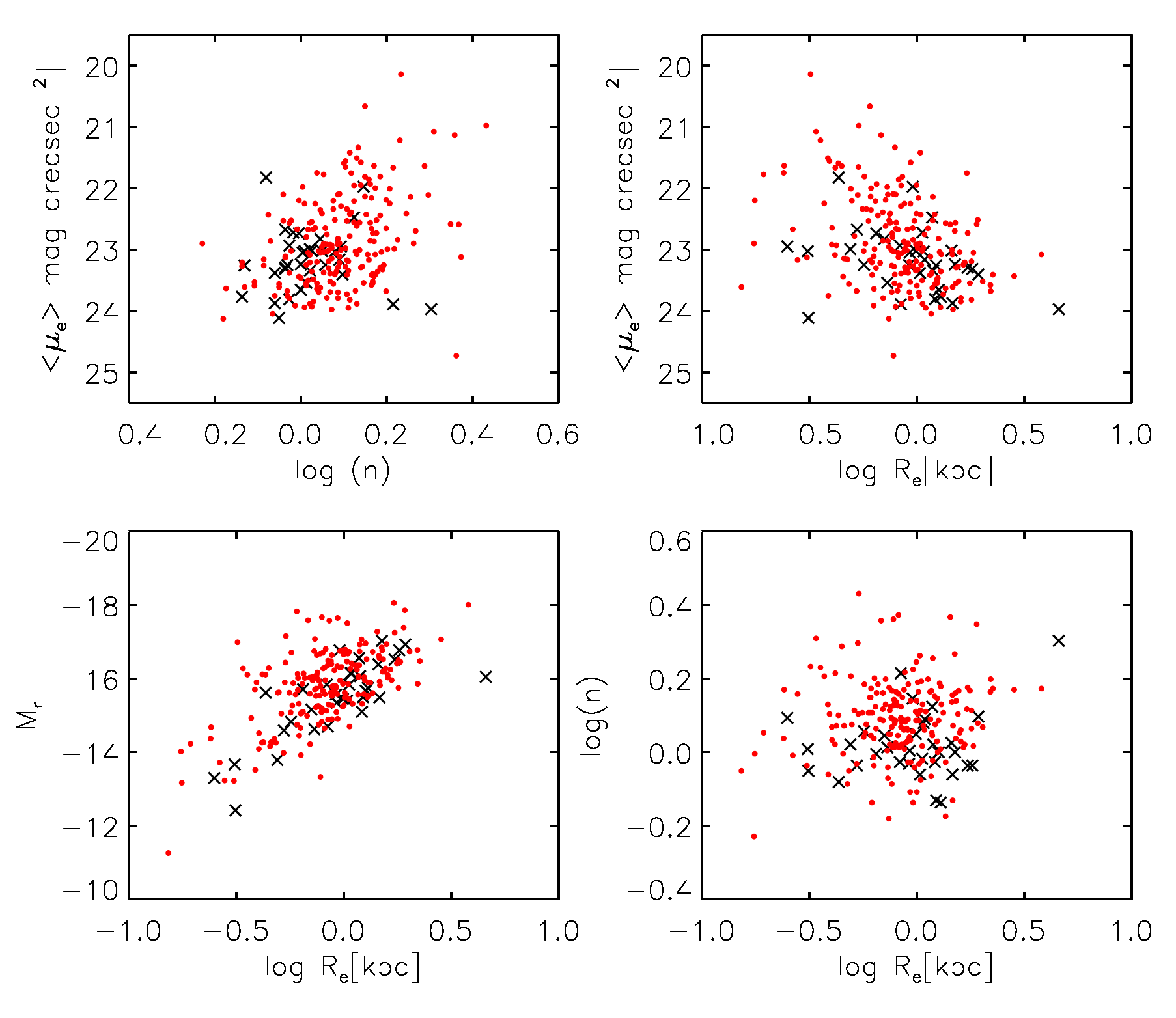}
\caption{Correlation of structural parameters of dE galaxies distinguished by
nucleation. The galaxies with nucleation are plotted as red filled circles and those without nucleation are designated by black crosses.
}
\label{fig5} 
\end{figure}

\begin{figure}
	\graphicspath{ {./figs/} }
	\centering
	\includegraphics[trim=0 0 0 0,width=\columnwidth]{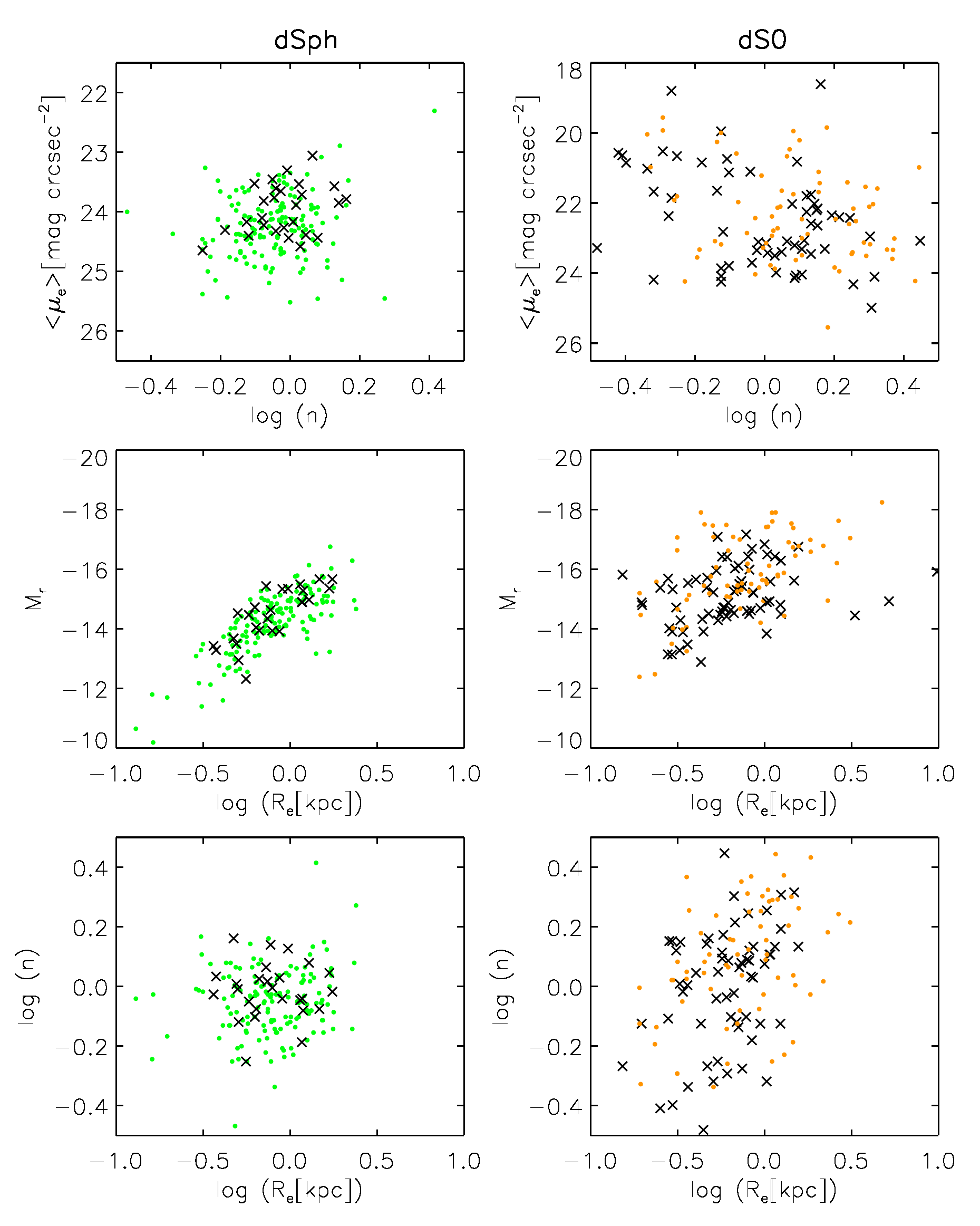}
	\caption{Correlation between structural parameters of dS0 and dSph galaxies distinguished by nucleation, dSph in the left column and dS0 in the right column. The galaxies with nucleation are plotted as coloured filled circles and those without nucleation are designated by black crosses.
	}
	\label{fig6}
\end{figure}

We examined the effect of nucleation on the correlations between structural parameters ($M_{r}$, n, $<\mu_{e}>$, $R_{e}$) for dS0, dE, and dSph galaxies and found that there is no significant  difference in most correlations. Figure \ref{fig5} (dE galaxies) and Figure \ref{fig6} (dS0 and dSph galaxies) show  correlations between structural parameters which seem to be of interest. As shown in the upper left panel of Figure \ref{fig5}, there is a weak correlation between $<\mu_{e}>$ and log (n) in nucleated dEs (denoted by red filled circles) with cc = -0.38, while there is no correlation in non-nucleated dEs (denoted by black crosses) with cc = 0.01. In contrast, there is no considerable difference in the correlation between $<\mu_{e}>$ and log ($R_{e}$) presented in the upper right panel of Figure \ref{fig5}. As can be seen in these figures, there is no non-nucleated dEs that have surface brightness brighter than $<\mu_{e}>   \approx 22$. Most high surface brightness dEs are due to strong nucleation. While the relation between $M_{r}$ and log ($R_{e}$) shows strongest   correlation among the correlations between structural parameters of dE galaxies, there is no appreciable difference between nucleated dEs and non-nucleated dEs. They show regression coefficients of $-3.02 \pm 0.25$ and $-3.25 \pm 0.46$, respectively. However, there is some difference in the luminosity as shown in the lower left panel of Figure \ref{fig5}. There are a lot of bright nucleated dEs compared with the mean relation between $M_{r}$ and log ($R_{e}$) of non-nucleated dEs. The difference in the luminosity distribution between nucleated dEs and non-nucleated dEs are more easily seen in the log (n) $vs$ log ($R_{e}$) diagram presented in the lower right panel of Figure \ref{fig5} which shows no correlation for both nucleated and non-nucleated dEs.

Figure \ref{fig6} shows the correlations between structural parameters for dS0 and dSph galaxies. For the correlation between $<\mu_{e}>$ and log (n), shown in the top panels, the effect of nucleation is somewhat pronounced in dSph galaxies but negligible in dS0 galaxies. The correlations show opposite trend in nucleated and non-nucleated dSphs although they show negligible (cc = 0.11) and weak (cc =-0.33) correlation respectively for nucleated and non-nucleated dSphs. As shown in the middle panels, contrary to the case for dE galaxies, there is a significant difference in the correlations between $M_{r}$ and log ($R_{e}$) for nucleated and non-nucleated dSphs because the regression coefficient of nucleated dSphs is $-2.7 \pm 0.26$ with confidence interval of (-3.96, -1.24) while that of non-nucleated dSphs is $-5.03 \pm 0.38$ with the confidence intervals of (-6.97, -4.53). Since they have cc = -0.65 for nucleated dSphs and cc = -0.94 for non-nucleated dSphs, The dS0 galaxies show a similarly significant difference in the correlation between $M_{r}$ and log ($R_{e}$). Their regression coefficients are $-3.45 \pm 0.38$ with confidence interval of (-5.12, -2.26) for nucleated dS0s and $-1.39 \pm 0.38$ with  with confidence interval of (-2.21, -0.43) for non-nucleated dS0s, respectively. But, there are differences in the regression coefficients between dSphs and dS0s. Steeper slope is found in the non-nucleated dSphs while steeper slope is found in the nucleated dS0s. For the correlations between log (n) and log ($R_{e}$), dSph galaxies show no difference between nucleated and non-nucleated dSphs whereas dS0 galaxies show a clear difference because the slopes are opposite. However, this difference is not statistically significant since the slope of nucleated dS0s is within the confidence interval of the slope of non-nucleated dS0s. 

The absence of a significant difference in most correlations for structural parameters between the nucleated and non-nucleated dwarfs is due to the small size of the nuclear region that is difficult to affect the global structure of galaxies. Moreover, we derived the structural parameters using two component models for these galaxies, one for the nucleus and the other for the main body of galaxies. Hence, the structural parameters are not much 
affected by the presence of the nucleus.

\subsection{Environmental Dependence of Structural Parameters} \label{subsec:environment}

We use the local galaxy density as a measure of environment. We derive the local galaxy density from the $n$th nearest neighbour method with $n=5$, 
$$\Sigma={n \over{4\pi r_{p,n}^2}} $$
where $r_{p,n}$ is the projected distance from the target galaxy to the $n$th nearest neighbour galaxy. Criteria of selecting neighbour galaxies are the same as those adopted by \citet{ann17}. That is, we use the two parameters, $M_{r}^{\ast}$ and $\Delta V^{\ast}$, to constrain the neighbour galaxies. The former parameter (limiting magnitude) is set to  $M_{r}^{\ast}=-15.2$ and the latter (linking velocity) is set to $\Delta V^{\ast}=1000 $km s$^{-1}$. We normalize the local galaxy density ($\Sigma$) by the mean local galaxy density ($\overline{\Sigma}$) derived for the galaxies in the local universe ($z\lesssim 0.01$) using SDSS DR7. 

\begin{figure}
	\graphicspath{ {./figs/} }
	\centering
	\includegraphics[trim=15 20 150 150,width=\columnwidth]{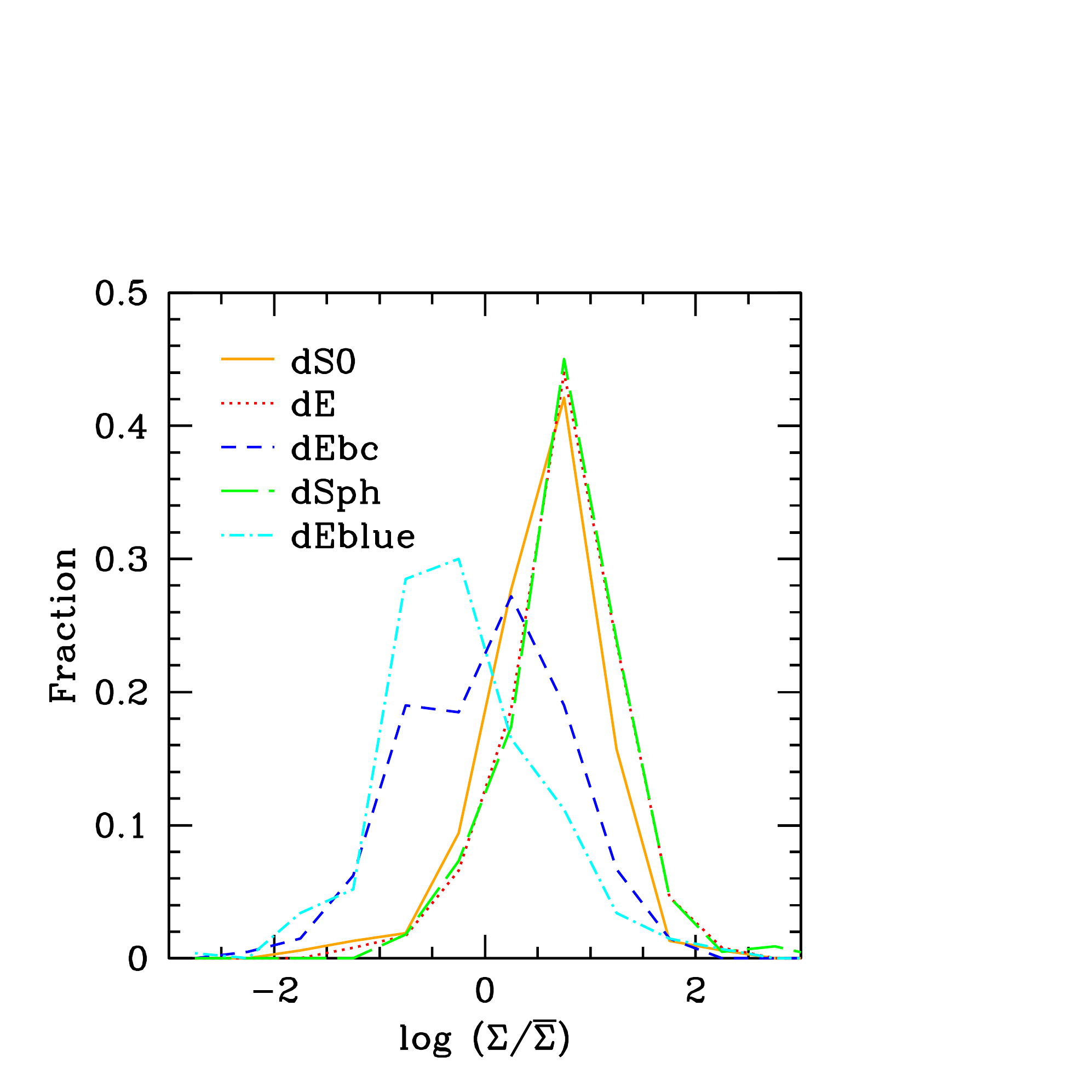}
	\caption{Local galaxy density distributions ($\Sigma/\overline{\Sigma}$) of dwarf elliptical-like galaxies. }
	\label{fig7}
\end{figure}

Figure \ref{fig7} shows the distribution of the local galaxy density of the five sub-types of dE-like galaxies. There is little difference among the three types of dS0, dE, and dSph which are located in the high density regions. The local galaxy densities of dE$_{bc}$ and dE$_{blue}$ galaxies are much different from those of dS0, dE and dSph galaxies with a larger difference in dE$_{blue}$ galaxies. The distribution of the local galaxy density of dE$_{bc}$ galaxies is similar to that of dE$_{blue}$ galaxies but it shows more fraction in high density regions than dE$_{blue}$. The low density tails of the three types (dS0, dE and dSph) are partly due to the galaxies in the survey boundary. However, some of them are thought to be genuinely extremely isolated galaxies.  

\begin{figure}
	\graphicspath{ {./figs/} }
	\centering
	\includegraphics[width=\columnwidth]{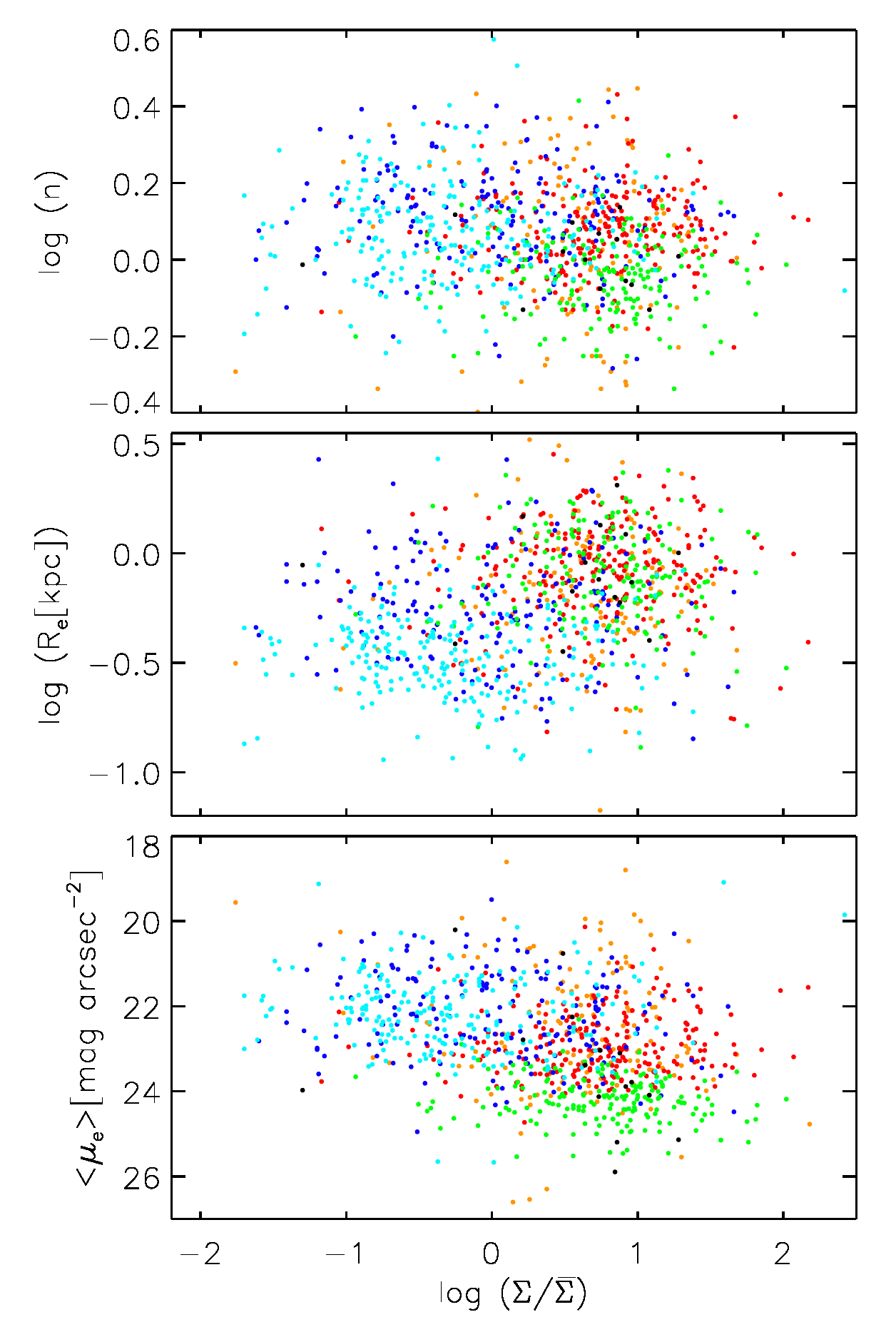}
	\caption{Local galaxy density ($\Sigma/\overline{\Sigma}$) versus three parameters n, $R_{e}$, and $<\mu_{e}>$. Colour codes for sub-types are the same as Figure 3. (top) S\'{e}rsic index n, (middle) effective radius $R_{e}$, (bottom) mean surface brightness wihtin $R_{e}$ $<\mu_{e}>$.
	}
	\label{fig8}
\end{figure}

The environmental dependence of the structural parameters ($n$, $R_{e}$, and $<\mu_{e}>$) of dE-like galaxies is explored using scatter diagrams as shown in Figure \ref{fig8}. 
There is not much difference in the local galaxy densities of dS0, dE, and dSph galaxies which are somewhat different from those of dE$_{bc}$ and dE$_{blue}$. As shown in Figure \ref{fig8}, the local galaxy density of each sub-type of dE-like galaxies spreads largely with the mean local galaxy densities of $0.56 \pm 0.55$, $0.74 \pm 0.57$, $0.03 \pm0.73$, $0.75 \pm 0.53$, and $-0.22 \pm 0.73$, for  dS0, dE, dE$_{bc}$, dSph, and dE$_{blue}$ galaxies, respectively.  Among the five sub-types, on average, dE and dSph galaxies are located in the highest density regions while dE$_{blue}$ galaxies are located in the lowest density regions. 
The non-negligible fraction of dE galaxies in the low density regions (log $\Sigma/\bar{\Sigma} < -1$) is quite puzzling because dE galaxies are mostly observed in high density regions. However, some of them are not genuinely isolated galaxies but are located in the survey boundary. There are only a few dEs that are located in the low density regions with log $\Sigma/\bar{\Sigma} < -1$.
The local galaxy density of dE$_{bc}$ galaxies overlaps with that of dE$_{blue}$ in the low-density regions and those of dS0, dE, and dSph galaxies in the high-density regions. The low-density environment of dE$_{blue}$ galaxies, similar to that of dwarf irregular (dI) galaxies reported in \citetalias{ann15}, along with their blue colours imply that they are dI galaxies with round shape. 
As a whole, there is no dependence of $R_{e}$ on the local galaxy density and a weak dependence of log (n) and $<\mu_{e}>$ on log ($\Sigma/\overline{\Sigma}$). That is, S\'{e}rsic index (n) decreases with increasing local galaxy density, and  $<\mu_{e}>$ increases, i.e., becoming low surface brightness, with increasing local galaxy density .

\begin{figure*}
	\graphicspath{ {./figs/} }
	\centering
	\includegraphics[trim=60 280 50 80,width=\textwidth]{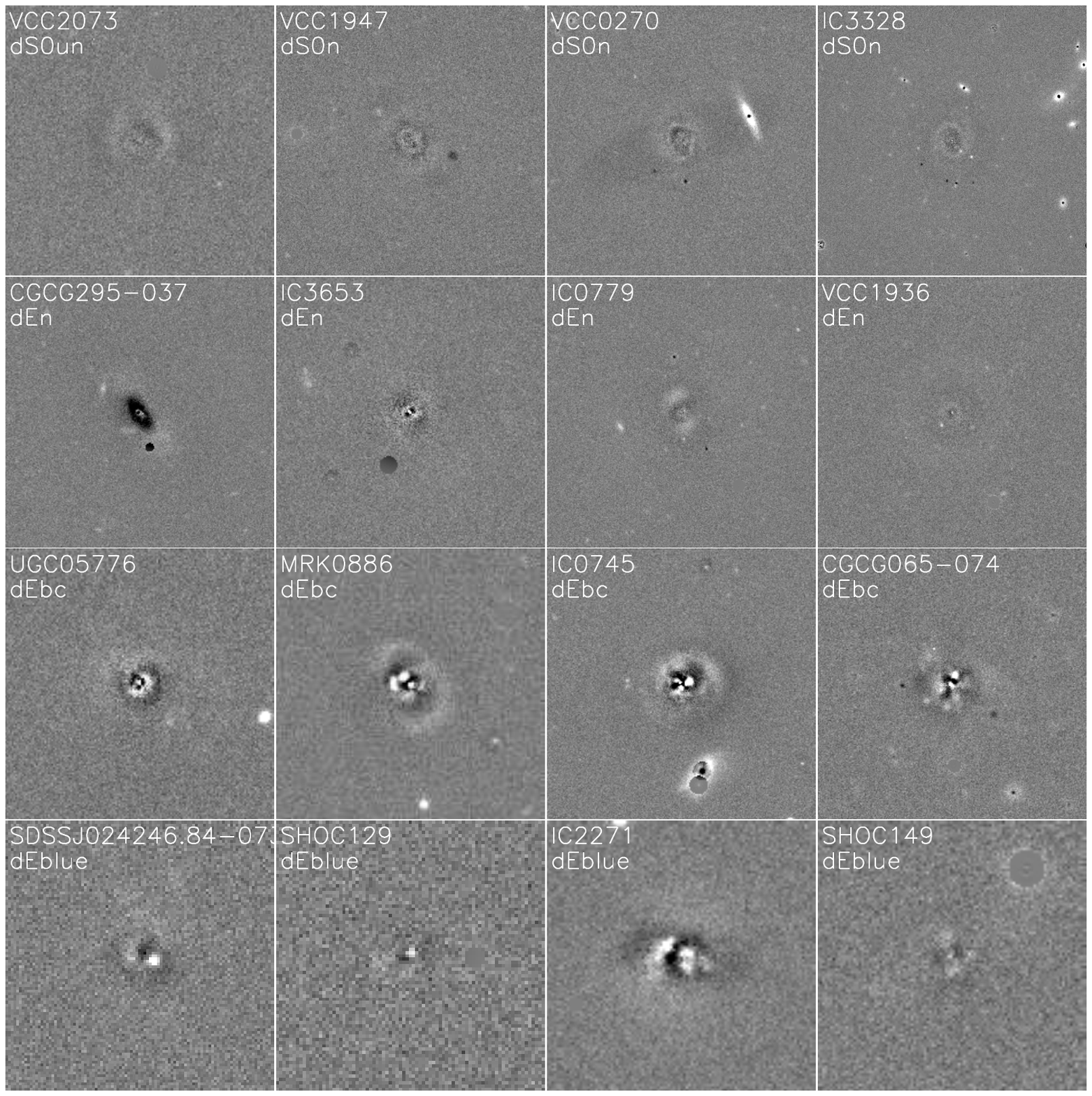}
	\caption{Sample of residual images of dwarf elliptical–like
		galaxies. (1$^{st}$ row) dS0 galaxies, (2$^{nd}$ row) dE galaxies,
		(3$^{rd}$ row) dE$_{bc}$ galaxies, (4$^{th}$ row) dE$_{blue}$ galaxies. }
	\label{fig9}
\end{figure*}

\subsection{Spiral features} \label{subsec:residual}

The most striking features observed in a large number of dE-like galaxies are the underlying spiral structures which are thought to be closely related to their origin. We investigate the residual images obtained by 
subtracting the GALFIT model images from the observed ones to see the underlying structures. 
Examples of residual images are given in Figure \ref{fig9} where we can see features such 
as spiral arm remnants, bars, and rings. We refer them as disk features below. We summarize 
the fractions of five sub-types of dE-like galaxies that show disk features in Table \ref{tab4} where the subscripts $'n'$ and $'un'$ denote the presence and absence of 
nucleation, respectively.

\begin{table*}
  \centering
  \caption{The fraction of disk features of dE-like galaxies. }
    \begin{tabular}{|c|c|c|c|c|c|c|c|c|c|c|c|}
    \hline
        $type$  & dS0 & dS0$_{un}$ & dS0$_{n}$  & dE  & dE$_{un}$  & dE$_{n}$  & dSph & dSph$_{un}$ & dSph$_{n}$ & dE$_{bc}$  & dE$_{blue}$ \\
    \hline
       fraction & 0.25  & 0.19  & 0.32  & 0.34  & 0.14  & 0.38  & 0.00  & 0.00  & 0.01  & 0.16  & 0.09  \\
    \hline
    \end{tabular}%
    \label{tab4}%
\end{table*}%

\begin{figure}
\graphicspath{ {./figs/} }
\centering
\includegraphics[trim=0 0 0 0,width=\columnwidth]{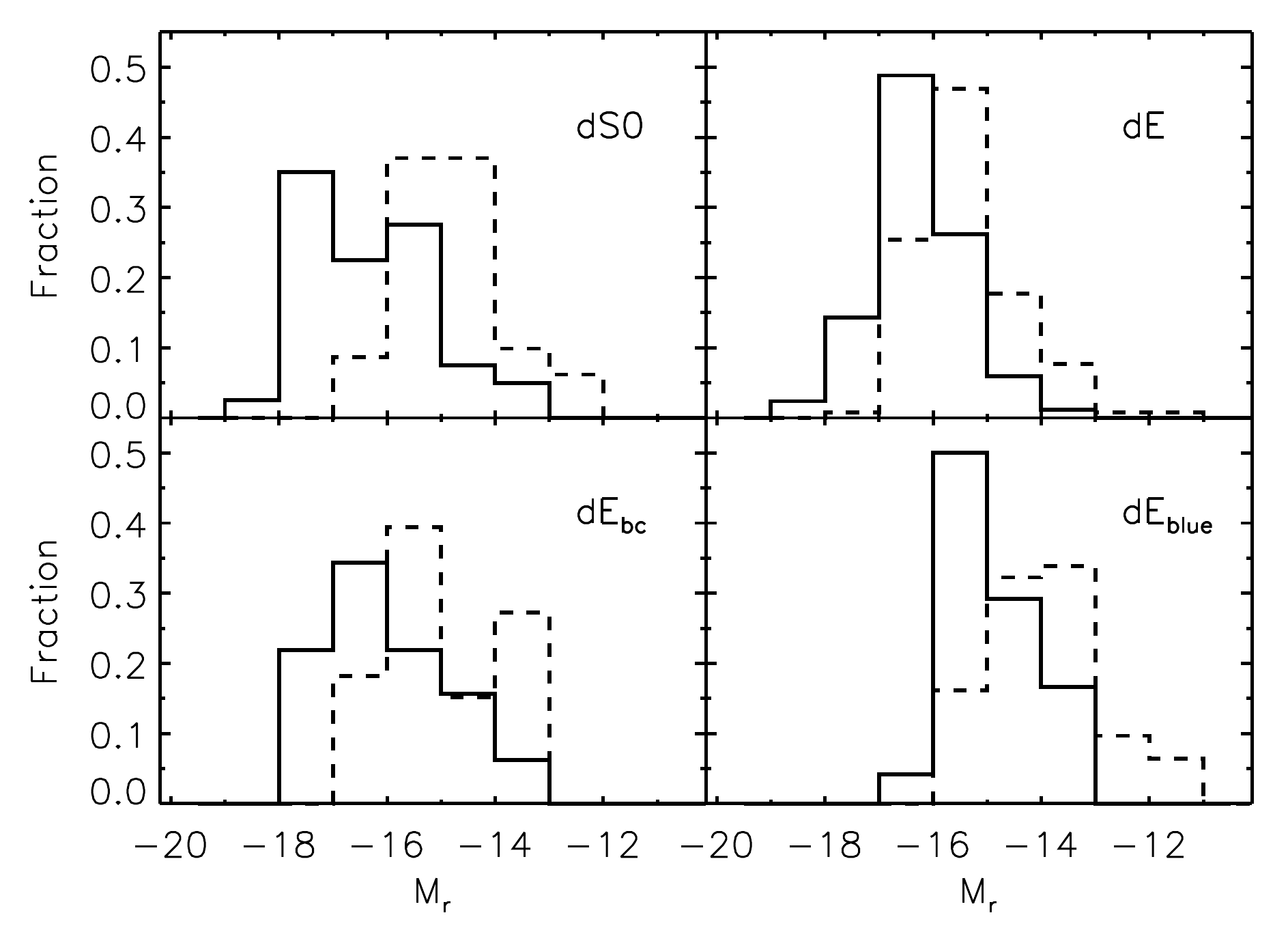}
\caption{Frequency distribution of dE-like galaxies with and without disk features as a function of $M_{r}$. Galaxies showing disk features are plotted by solid lines and those showing no disk features are plotted by dashed lines. (upper left) dS0 galaxies, (upper right) dE galaxies, (bottom left) dE$_{bc}$ galaxies, (bottom right) dE$_{blue}$ galaxies
}
\label{fig10}{\Large {\Large }}
\end{figure}

As shown in Table \ref{tab4}, about $17\%$ of our sample of dE-like galaxies shows disk features 
in the residual images. However, the disk features we found are mainly observed in dS0 and dE galaxies, and the residual features of dE$_{bc}$ and dE$_{blue}$ galaxies are thought to be not spiral structures but irregular star forming regions. They are sometimes double or multiple core-like structures of enhanced star formation (see Figure \ref{fig9}). Therefore, if we consider spiral features in dS0, dE, and dSph galaxies only, it is increased to $\sim20\%$ of these galaxies. This fraction is much larger than the results of \citet{lis06} who reported $10\%$ in the Virgo cluster, and a little smaller than $\sim20-25\%$ fraction found in early-type dwarf galaxies in the Virgo and the Coma clusters \citep{agu16}.

The highest frequency of spiral features ($\sim34\%$) appears in dE galaxies. The dE with nucleation (dE$_{n}$) have spiral features in $38\%$. The difference in the fraction of spiral features between galaxies with and without nucleation is also pronounced in dE galaxies, $14\%$ in dE$_{un}$ and $38\%$ in dE$_{n}$. However, the most interest founding is the absence of disk features in dSph galaxies. If we apply a single component model to the luminosity distribution of dSph galaxies that have  nucleation, some dSph galaxies show small features in residual images, but there is no disk features seen in the residual images obtained from the two component models of which one assumes a nuclear component.

Figure \ref{fig10} shows the fractional frequency distributions of dE-like galaxies as a function of $M_{r}$, divided into galaxies with and without disk features. The galaxies with disk features are represented by solid lines and those with no disk features are plotted by dashed lines. We except dSph galaxies because they do not show disk features. For most dE-like galaxies, disk features are more likely to be found in bright galaxies. This trend is most pronounced in dS0 galaxies of which disk features are mostly spiral arm remnants. There are only a small fraction of dS0 galaxies brighter than $M_{r}=-16$ that do not have disk features. Similar trend is found for dE and dE$_{bc}$ galaxies that are brighter than $M_{r}=-17$. In dE$_{blue}$ galaxies, the frequency distribution of disk features appears to be symmetrical in the sense that more galaxies with disk features at the bright part and more galaxies without disk features at the faint part. 

\begin{figure}
\graphicspath{ {./figs/} }
\centering
\includegraphics[trim=0 0 0 0,width=\columnwidth]{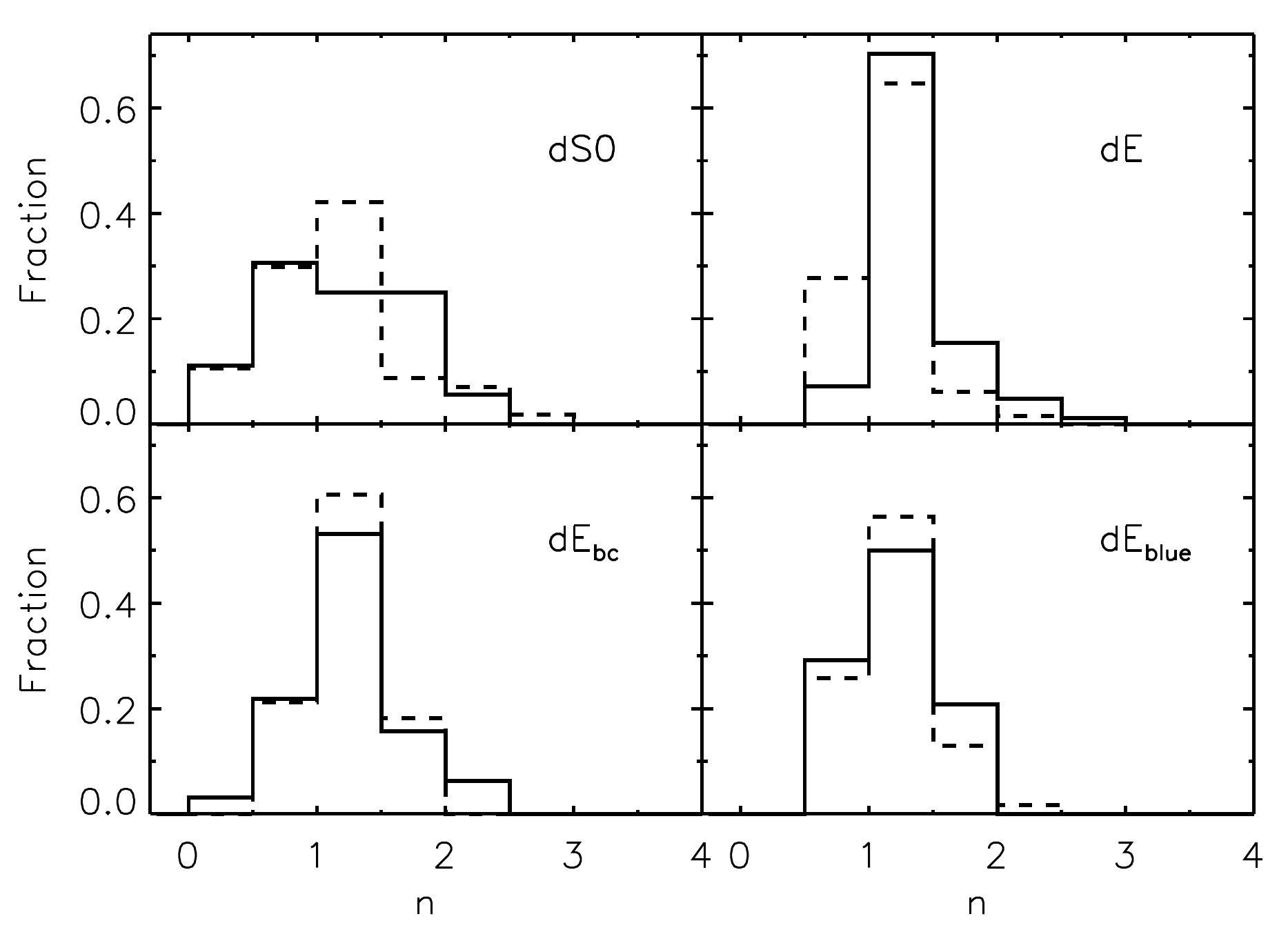}
\caption{Frequency distribution of dE-like galaxies with and without disk features as a function of S\'{e}rsic index, n. 
Galaxies showing disk features are plotted by solid lines and those showing no disk features are plotted by dashed 
lines. (upper left) dS0 galaxies, (upper right) dE galaxies, (bottom left) dE$_{bc}$ galaxies, (bottom right) dE$_{blue}$ galaxies
}
\label{fig11}
\end{figure}

Figure \ref{fig11} shows the frequency distributions of S\'{e}rsic index in dE-like galaxies. We represent the galaxies with disk features by solid lines and those without disk features by dashed lines. In general, there is no significant difference between the galaxies with disk features and those without disk features. However, there is a weak tendency of a larger n for galaxies with disk features in dS0 and dE galaxies.

\begin{figure}
	\graphicspath{ {./figs/} }
	\centering
	\includegraphics[trim=0 0 0 0,width=\columnwidth]{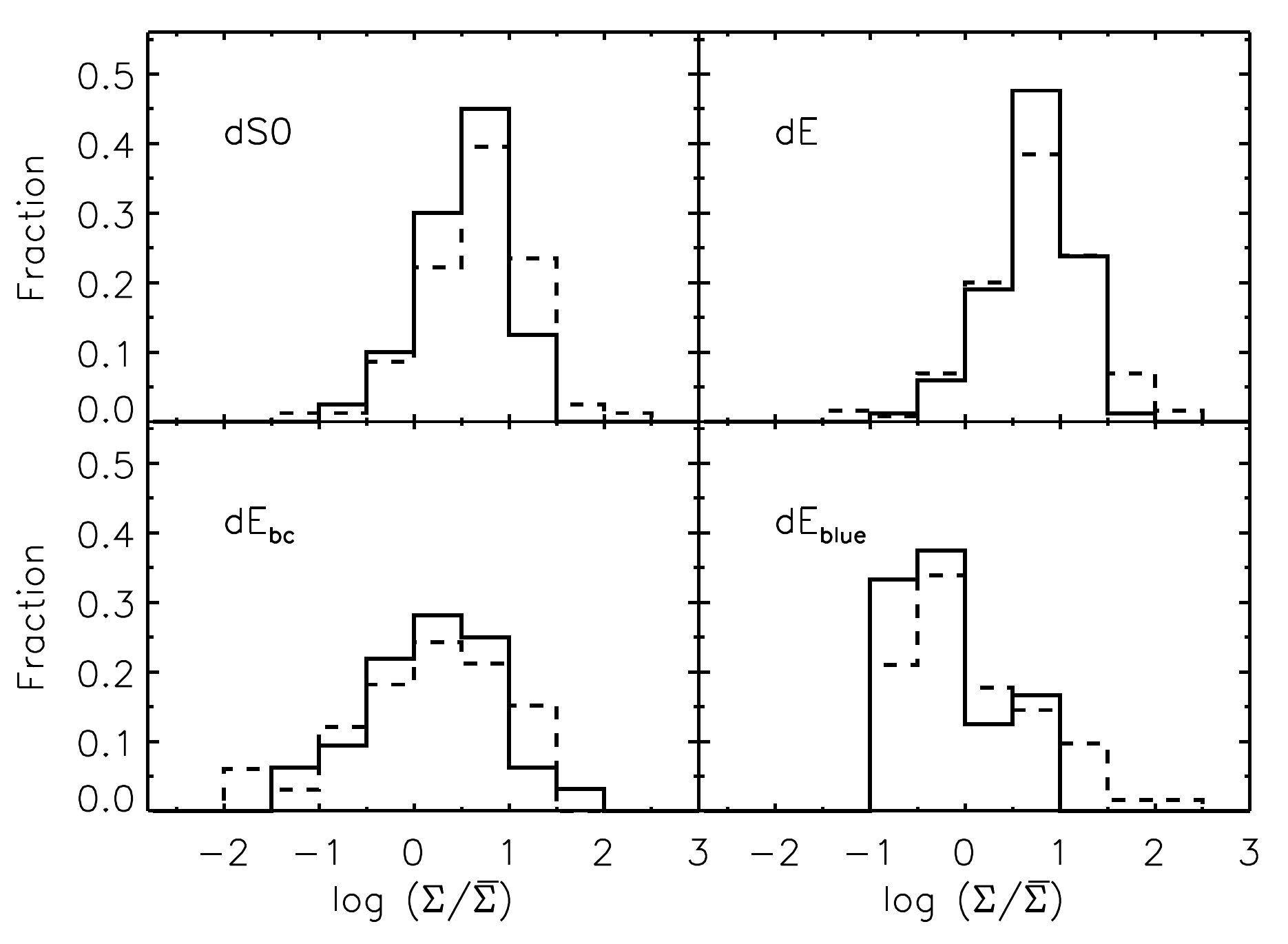}
	\caption{Frequency distribution of dE-like galaxies with and without disk features as a function of the local background density. 
		Galaxies showing disk features are plotted by solid lines and those showing no disk features are plotted by dashed 
		lines. (upper left) dS0 galaxies, (upper right) dE galaxie{\large }s, (bottom left) dE$_{bc}$ galaxies, (bottom right) dE$_{blue}$ galaxies
	}
	\label{fig12}
\end{figure}

\subsection{Environmental Dependence of Spiral Feature} \label{subsec:environment}

The environmental dependence of the frequency distributions of disk features in dE-like galaxies is examined in Figure \ref{fig12}. As can be seen in Figure \ref{fig12}, there is no significant difference in the frequency distributions of the fraction of galaxies with disk features and those with no disk features except for dE$_{blue}$ galaxies which are likely to be located in the under-dense regions. However, there is some difference in the distribution of the local galaxy density for galaxies with and without disk features. The dS0 and dE galaxies with no disk features are more likely to be found in the high density regions. In particular, there is no dS0 galaxies with disk features in the local galaxy density higher than log ($\Sigma/\bar{\Sigma}) \sim 1.5$. The dE galaxies show environmental dependence similar to dS0 galaxies but a negligible fraction of dE galaxies with disk features shows background density higher than log ($\Sigma/\bar{\Sigma}) \sim 1.5$. 

Among the dE-like galaxies, dE$_{bc}$ galaxies show the broadest distribution of the local galaxy density but no significant difference between galaxies with disk features and those without disk features. Another feature worth to be noted is that a significant fraction of dE$_{bc}$ galaxies are located in the low density regions with log ($\Sigma/\bar{\Sigma}) < -1$. For dE$_{blue}$ galaxies, they are prevalent at under-dense regions, regardless of the presence/absence of disk features. On the contrary, at high density regions of log ($\Sigma/\overline{\Sigma}) > 1$, dE$_{blue}$ galaxies with no disk features are much more frequent than dE$_{blue}$ with disk features. It seems to be related to the properties of the disk features in dE$_{blue}$ galaxies. While the disk features in dS0 and dE galaxies are thought to be spiral arms, bars, and rings, which are leftover material after transformation from the late type disk galaxies to dS0 and dE galaxies, those observed in dE$_{blue}$ galaxies are sites of on-going active star formation. Since star formation is more active in the under-dense regions at the present time, dE$_{blue}$ galaxies with disk features are more likely to be found in the under-dense regions as shown in the bottom right panel of Figure \ref{fig12}.

\section{Discussion} \label{sec:disscusion}

The terminology of dE-like galaxies used in the present study includes five sub-types of dwarf galaxies; dS0, dE, dSph, dE$_{bc}$, and dE$_{blue}$. Among them, the first three sub-types are frequently found in the literature to represent the early-type dwarf galaxies, whereas the others are new ones introduced by  \citetalias{ann15} to represent blue-cored dwarf elliptical galaxy (dE$_{bc}$) and globally blue dwarf galaxy with round shape (dE$_{blue}$). \citet{ann17} considered dE$_{bc}$ galaxies as early-type dwarfs and dE$_{blue}$ as late-type dwarfs because of their active star formation like dwarf irregular (dI) galaxies. Actually, there is no difference between dE$_{blue}$ and dI except for the roundness. They are very similar to HII region-like BCDs with somewhat reduced star formation. The sub-types of dE-like galaxies are reported in some previous studies \citep{lis07, km13} but with different symbols.  In particular, the classification system of dwarf galaxies by \citet{km13} includes all the sub-types of \citetalias{ann15} if we consider dE$_{blue}$ as HII region-like BCD. Thus, the sub-types of dE-like galaxies such as those employed in the present study is useful for a detailed analysis of the structural properties of dwarfs although there is no consensus on the symbols.

There is a considerable difference between the structural parameters of the five sub-types of dE-like galaxies. In particular the S\'{e}rsic index n of dSphs is clearly distinguished from that of dEs when we apply two-component model of which one component is the S\'{e}rsic function to represent the main body of a galaxy and the other component represent the nuclear light excess due to nucleation. The K-S test of the distribution of $n$ for the two sub-types are statistically different with  $p < 0.05$ for most cases.The difference in the S\'{e}rsic index between dSphs and dEs reflects the morphological difference of the two sub-types. The morphology of dSphs is characterized by the lower surface brightness and shallower gradient in the luminosity distribution than dEs. The presence of nucleation does little affect the S\'{e}rsic index n fitted to the main body of dSphs and dEs. In contrast, there is no significant difference in n between dSphs and dEs if we neglect the nuclear component. Thus, proper treatment of the nuclear component is critical for a better understanding of the structure of early-type dwarfs since most early-type dwarfs show nucleation. 

The noticeable difference in some structural parameters among sub-types of dE-like galaxies also supports the usefulness of sub-types in the classification of dwarf galaxies. For example. there is a pronounced difference in the mean surface brightness within the effective radius, $<\mu_{e}>$ between dSph galaxies and others. The dSph galaxies have $<\mu_{e}> = 24.2\pm0.8$ while other types have $<\mu_{e}>  \lesssim23.0$ with the highest one in dE$_{blue}$ as $<\mu_{e}> = 22.2$. The small S\'{e}rsic index of dSph galaxies is closely related to the low surface brightness and the shallow gradient in the surface brightness. Although the distinction of dSph galaxies from dE galaxies has been made clearly for the galaxies in the LG, many previous studies (e.g., \citealt{kor85}) did not distinguish dSph galaxies from dE galaxies. However, dSph galaxies have structural properties different from those of dE galaxies besides the photometric properties. They are supposed to be pressure supported \citep{walk09,sal12} while dE galaxies are likely to be rotation supported \citep{ped02,simPre02,geh03,vanzee04,rij05,chi09,geh10,tol15,pen16}. They are dark matter dominated galaxies with $M/L > 100$ \citep{gal94,mat94,pry94,pry96,ger94,ols98,bat13,sim21}.

The lack of disk features in dSph galaxies, which is a unique property of dSph galaxies, provides a further observational evidence which make them
different from dE galaxies. The deficit of disk features in faint early-type dwarf galaxies of the Virgo cluster was noticed by \citet{lis06} who found that $> 50\%$ of early-type dwarfs at bright end have disk features while only a few percent of them have disk
features at $m_{B} >16$ which corresponds to $M_{B}\approx-15$. Since the
majority of faint early-dwarf galaxies in the VCC are supposed to be dSph galaxies
in \citetalias{ann15}, the lack of disk features in dSph galaxies is in a good agreement with \citet{lis06}.

Any scenarios for the origin of dSph galaxies are required to explain the aforementioned properties of dSph galaxies. In a broad category, there are two scenarios for the origin of dSph galaxies, of course dE-like galaxies in general. One is that they are primordial objects, i.e., they are the descendants of building blocks of the early universe. The other is that they are transformed objects from the late-type galaxies by the environmental effects \citep{lis07, jan14} such as ram pressure stripping \citep{gg72} and galaxy harassment \citep{moo98, moo99}. While ram pressure stripping of the cold gas in the late-type galaxies leads to shutdown of star formation, galaxy harassment thickens the disk of a galaxy by tidal heating \citep{to92} or tidal stripping. The small mass of dSph galaxies facilitates gas removal in these galaxies whether they are primordial objects or transformed ones. 
Since the mean stellar age of dSph galaxies which are originated from the primordial objects is thought to be much older than that of the dSph galaxies
transformed from the late-type galaxies falling into cluster environment, analysis of stellar age is required to clarify their origin. We will present a detailed
analysis of the SDSS spectra of the dE-like galaxies in the forthcoming papers.

The majority of early-type dE-like galaxies (dS0, dE, dSphs and dE$_{bc}$) are
supposed to be transformed from late-type galaxies because most of
the dE-like galaxies are located in group and cluster environment where
environmental effects such as ram pressure stripping and galaxy harassment are
expected to be operating. On the other hand, the primordial origin of the
dE-like galaxies is also plausible because at least $\sim5\%$ of them are
isolated galaxies which do not suffer environmental effects to transform
their morphology. \citet{ann17} also showed that some dE and
dSph galaxies which are satellites of isolated satellite systems are likely
to be primordial objects because they are located in the outer part of the
satellite systems where environmental effect is not effective. 
Thus, as noted by  \citet{lis09}, multiple origins of dE-like galaxies are very likely.

\section{Summary and conclusions}

We examined the luminosity distributions and spiral features of the five sub-types of dE-like galaxies in the local universe ($z \lesssim0.01$) by applying GALFIT to the $r-$band SDSS images. We derived structural parameters such as the effective radius ($R_{e}$), the mean surface brightness within $R_{e}$ ($<\mu_{e}>$), axial ratio ($b/a$) along with the S\'{e}rsic index ($n$). The structural parameters of the dE-like galaxies overlap a lot among sub-types but the most likely structural parameters of each sub-type could be distinguished from others. In particular, $<\mu_{e}>$ seems to be effective to distinguish dSph galaxies from other sub-types.

The distribution of the S\'{e}rsic index n of dE-like galaxies is much different from pure disk galaxies as well as elliptical galaxies.
More than $\sim90\%$ of dE-like galaxies have the S\'{e}rsic index between 0.5 and 2.0 except for the dE$_{bc}$ galaxies of which $\sim36\%$ of them have
the S\'{e}rsic index larger than n=2.
However, the detailed distribution of n depends on the sub-types. The dS0, dE, and dE$_{blue}$ have the largest fractions at n=1.0 $\sim 1.5$ while dSph peaks at n=0.5 $\sim 1.0$ and dE$_{bc}$ peaks at n=1.5 $\sim 2.0$.
It is crucial to add a nuclear component to fit the luminosity distribution of dE-like galaxies because most dE-like galaxies show nucleation which leads to a larger n if we apply a single-component model to fit the observed luminosity distribution.  
The nuclear component is approximated by either King profile or Nuker profile. 
The K-S test for the distribution of the S\'{e}rsic index between sub-types shows that they are statistically different populations. The distinction between sub-types, in particular between dSph and dE, may cast some insight into their origins.  

There are clear correlations between structural parameters, which depend on the sub-types.
For example, the Pearson correlation coefficient ($cc$) for the relation between  $<\mu_{e}>$ and log ($R_{e}$) is $0.46$. The structural parameters also correlate with the galaxy luminosity in the sense that bright galaxies have larger $R_{e}$ and higher $<\mu_{e}>$ than faint galaxies. The relation between the luminosity ($M_{r}$) and the mean surface brightness ($<\mu_{e}>$) of the dE-like galaxies shows the same trend with the relation between the luminosity and the central
surface brightness of early-type dwarf galaxies, that is the surface brightness of early-type dwarf galaxies, Sph in Kormendy's terminology, increases, i.e., becoming brighter, with increasing luminosity. This trend is opposite to the trend observed in the early-type giant galaxies \citep{kor85}. 

A significant fraction of dE-like galaxies show disk features such as spiral arm, bar, lens, and ring structures but it depends on sub-types. The two sub-types, dS0, and dE, have disk features in $\sim30\%$ of them, whereas there is virtually no dSph galaxies that have disk features. 
The residual images of dE$_{bc}$ and dE$_{blue}$ galaxies show somewhat different features from those of dE and dS0. Especially, dE$_{blue}$ galaxies show the star forming regions of irregular
shapes rather than the spiral features. The dE$_{bc}$ galaxies show mixed
features. It is apparent that spiral arms are more frequent in the bright
dS0 and dE galaxies but there is no correlation with the S\'{e}rsic index.
The disk features are more likely to be found in the low density regions than the high density regions, however, the dependence is very weak.
The presence and absence of nucleation do not affect the disk features
of dS0 galaxies but  more frequent disk features in the nucleated dE
galaxies. On the other hand, disk features depend strongly on the
galaxy luminosity. The fraction of disk features is largest at the bright end
of dE-like galaxies. The absence of disk features in dSph galaxies is in
a good agreement with the luminosity dependence{\large } of disk features.

It seems likely that the majority of the three sub-types of dE-like
galaxies (dS0, dE, and dE$_{bc}$) are transformed from the late-type galaxies
in groups and clusters but there is other route for the formation of these
galaxies.

\section*{Acknowledgments}
This work was supported partially by the NRF Research grant 2015R1D1A1A09057394.


\section*{Data Availability}
The original data underlying this article are available in SDSS DR7.
And  additional data are available upon request.



\bibliography{seoannRev1} 
\bibliographystyle{mnras}






\appendix

\section{Some extra material}
Table \ref{tab:a1} summarizes the correlation analysis applied to the data show in Figure \ref{fig4}. The columns give the following information. We applied linear least-squares fittings (Y=  a + bX) and calculated the Pearson correlation coefficients and the confidence interval for the slope from bootstrap resampling. 

\begin{table*}
\caption{The correlation analysis of the structural parameters.}
\label{tab:a1}
\begin{tabular}{llccccc}
  \hline
Parameters     & Galaxy    & a         & b          & Standard  & Pearson    & Confidence    \\ 
                       &  type       &(intercept) &(slop)     &  deviation    & coefficient  & interval \\
  \hline
	                & dS0    & -15.93±0.12 & -2.74±0.29 & 1.20 & -0.63 & (-3.89,-1.65)         \\
                         & dE     & -15.96±0.06 & -3.00±0.22 & 0.82 & -0.67 & (-3.66,-2.40)      \\
   $M_{r}$ vs. log (R$_{e}$)   & dE$_{bc}$   & -16.56±0.10 & -2.88±0.29 & 0.98 & -0.59 & (-3.45,-2.34)         \\
                           & dSph   & -14.52±0.07 & -3.18±0.23 & 0.86 & -0.71 & (-4.53,-1.85)         \\
                          & dE$_{blue}$ & -15.92±0.13 & -3.09±0.24 & 0.93 & -0.65 & (-4.25,-1.85)     \\
  \hline
			& dS0    & 21.58±0.26  & 0.65±0.19  & 1.54 & 0.29  & (0.33,1.05)        \\
                          & dE     & 23.89±0.18  & -0.79±0.14 & 0.67 & -0.35 & (-1.16,-0.40)       \\
 $<\mu_{e}>$ vs. log (n)     & dE$_{bc}$   & 23.63±0.26  & -1.11±0.18 & 1.02 & -0.41 & (-1.44,-0.80)        \\
                          & dSph   & 23.73±0.14  & 0.53±0.13  & 0.64 & 0.29  & (-0.12,1.86)          \\
                          & dE$_{blue}$ & 22.42±0.19  & -0.20±0.14 & 1.12 & -0.09 & (-0.78,0.47)          \\
 \hline
    			 & dS0    & 27.75±1.29  & 0.35±0.08  & 1.51 & 0.34  & (0.15,0.48)         \\
                           & dE     & 26.84±0.63  & 0.25±0.04  & 0.67 & 0.38  & (0.12,0.35)     \\
 $<\mu_{e}>$ vs. $M_{r}$        & dE$_{bc}$   & 26.07±1.04  & 0.25±0.07  & 1.08 & 0.27  & (0.12,0.38)      \\
                           & dSph   & 25.7±0.57   & 0.10±0.04  & 0.66 & 0.19  & (0.01,0.16)         \\
                           & dE$_{blue}$ & 25.55±0.86  & 0.23±0.06  & 1.09 & 0.25  & (-0.09,0.45)         \\
 \hline
 			& dS0    & 22.67±0.14  & 1.67±0.36  & 1.49 & 0.37  & (0.47,2.65)        \\
                           & dE     & 22.99±0.05  & 1.06±0.18  & 0.67 & 0.36  & (0.63,1.47)       \\
$<\mu_{e}>$ vs. log (R$_{e}$)   & dE$_{bc}$   & 22.67±0.10  & 2.14±0.29  & 0.99 & 0.47  & (1.59,2.65)         \\
                         & dSph   & 24.34±0.05  & 1.04±0.16  & 0.60 & 0.43  & (0.19,1.84)        \\
                         & dE$_{blue}$ & 22.98±0.14  & 1.75±0.27  & 1.03 & 0.40  & (1.20,2.45)      \\ 
 \hline
\end{tabular}
\end{table*}





\bsp	
\label{lastpage}
\end{document}